\documentclass[a4paper,11pt]{article}
\usepackage{jcappub}
\usepackage{lineno}
\usepackage{aas_macros}
\usepackage{mathtools}
\usepackage{siunitx} 

\hypersetup{ colorlinks=true, linkcolor=blue, urlcolor=blue, citecolor=blue, filecolor=blue }
\usepackage[svgnames]{xcolor}
\usepackage[normalem]{ulem}

\usepackage[nameinlink,noabbrev]{cleveref}
\Crefname{equation}{Eq.}{Eqs.}
\Crefname{section}{Sect.}{Sects.}
\Crefname{figure}{Fig.}{Figs.}
\crefname{equation}{Equation}{Equations}
\crefname{section}{Section}{Sections}
\crefname{figure}{Figure}{Figures}
\creflabelformat{equation}{#2#1#3}
\newcommand{\nCref}[1]{\Cref{#1}}

\usepackage{multirow}
\usepackage{booktabs}

\arxivnumber{2410.21457} 
\title{A revisited Correction to the Halo Mass Function for local-type Primordial non-Gaussianity}

\author[1]{L. Fiorino}
\author[2]{, S. Contarini}
\author[1,3,4]{, F. Marulli}
\author[2]{, A. G. Sánchez}
\author[1,3,4]{, M. Baldi}
\author[2]{, A. Fiorilli}
\author[1,3,4]{, L. Moscardini}
\affiliation[1]{Dipartimento di Fisica e Astronomia ``Augusto Righi" - Alma Mater Studiorum Università di Bologna, via Piero Gobetti 93/2,
40129 Bologna, Italy}
\affiliation[2]{Max Planck Institute for Extraterrestrial Physics, Giessenbachstrasse 1, 85748 Garching, Germany}
\affiliation[3]{INAF-Osservatorio di Astrofisica e Scienza dello Spazio di
Bologna, Via Piero Gobetti 93/3, 40129 Bologna, Italy}
\affiliation[4]{INFN-Sezione di Bologna, Viale Berti Pichat 6/2, 40127 Bologna, Italy}

\emailAdd{luca.fiorino2@studio.unibo.it}

\abstract{We investigate the effect of primordial non-Gaussianities on halo number counts using N-body simulations with different values of $f_{\rm NL}^{\rm loc}$. We show how current theoretical models fail to adequately describe the non-Gaussian mass function of halos identified with different overdensity thresholds, $\Delta_{\rm b}$. We explain how these discrepancies are related to a variation in the density profile of dark matter halos, finding that the internal steepness (i.e. the compactness) of halos depends on the value of $f_\mathrm{NL}^\mathrm{loc}$. We then parametrize these deviations in halo number counts with a factor $\kappa(\Delta_\mathrm{b})$ that modifies the linear density threshold for collapse according to the halo identification threshold used, defined with respect to the Universe background density. We rely on a second-degree polynomial to describe $\kappa$ and employ a Bayesian analysis to determine the coefficients of this polynomial. In addition, we verify the independence of the latter on the sign and absolute value of $f_\mathrm{NL}^\mathrm{loc}$. Finally, we show how this re-parametrization prevents the extraction of biased constraints on $f_\mathrm{NL}^\mathrm{loc}$, correcting for large systematic errors especially in the case of halos identified with high density thresholds. This improvement is crucial in the perspective of deriving cosmological constraints with the non-Gaussian mass function from real data, as different mass definitions can be employed depending on the properties of the survey.}

\begin{document}
\maketitle
\flushbottom

\section{Introduction}\label{sec:intro}
The theory of inflation \cite{Guth1982, Hawking1993} assumes that the early Universe has experienced a brief period of rapid expansion, providing a compelling explanation for the emergence of initial density perturbations, which eventually gave rise to the large-scale structures (LSS) observed today. Consequently, inflation has become a cornerstone of modern cosmology. However, since a direct observation of inflation is not possible, the mechanisms behind it remain unclear. As a result, cosmologists focus on identifying observable signatures that can help distinguish between various inflationary models \cite{Achucarro2022}.

Primordial non-Gaussianities (PNG) stand out as one of the most promising observational signatures for probing the inflationary process, as nearly all inflationary models predict their existence \cite{Liguori2003, Seery2005_I, Seery2005_II}. In the context of PNG, the distribution of primordial perturbations, which seed the formation of LSS, is assumed to deviate from a Gaussian distribution. These deviations are quantified by the parameter $f_\mathrm{NL}^\mathrm{X}$, where ``X'' denotes different shapes corresponding to various inflationary scenarios, each leaving distinct imprints on the primordial bispectrum \cite{Takahashi2014}. Among the various shapes of PNG, such as orthogonal and equilateral ones, the local type of PNG --- described by the parameter $f_\mathrm{NL}^\mathrm{loc}$ --- has garnered particular attention as the consistency relations of the single field slow roll inflation \cite{Maldacena2003, Acquaviva2003, Creminelli2004} predict $f_\mathrm{NL}^\mathrm{loc}$ of order $10^{-2}$. Therefore, a detection of $f_\mathrm{NL}^\mathrm{loc}\gg10^{-2}$ can rule out single-field inflation, favoring multi-field models that instead predict $f_\mathrm{NL}^\mathrm{loc}\sim1$ \cite{Bartolo2004, DePutter2017_II}. 

During the past two decades, Cosmic Microwave Background (CMB) experiments such as the \emph{Wilkinson Microwave Anisotropy Probe} (WMAP) \cite{WMAP2009, WMAP2011, WMAP2013} and \emph{Planck} \cite{Planck2014_XXIV, Planck2016_XVII, Planck2020_IX} have significantly constrained the magnitude of this form of PNG. In this regard, the latest \emph{Planck} results indicate $f_\mathrm{NL}^\mathrm{loc}=-0.9\pm5.1$. However, as the CMB's ability to provide further information is nearly exhausted, future galaxy surveys, such as the \emph{Dark Energy Spectroscopic Instrument} (DESI) \cite{DESI2016} and the \emph{Euclid} mission \cite{Blanchard2020, Mellier2024}, are expected to surpass the CMB in constraining power for PNG \cite{amendola2018}. The latter are indeed expected to impact the distribution of LSS and the current best constraints on $f_\mathrm{NL}^\mathrm{loc}$ comes from the modeling of the 3D power spectrum. For example, recent works analyzed the sample of \emph{Baryon Oscillation Spectroscopic Survey} (BOSS) luminous red galaxies (LRGs) \cite{Damico2022, Cabass2022} and the extended BOSS quasars
\cite{Castorina2019, Muller2022}, providing $\sigma(f_\mathrm{NL}^\mathrm{loc})\sim29$ and $\sigma(f_\mathrm{NL}^\mathrm{loc})\sim21$ respectively. However, future LSS observations can potentially reach $\sigma(f_\mathrm{NL}^\mathrm{loc})<1$ \cite{Seljak2009, Dore2014, Karagiannis2018, Ferraro2019}, allowing us to distinguish between different inflationary scenarios.

In terms of observational signatures, local-type PNG is expected to generate a scale-dependent bias proportional to $k^{-2}$, which leads to an enhancement of the amplitude of clustering on large scales in biased tracers of the matter distribution \cite{Matarrese2008, Dalal2008}, such as dark matter (DM) halos.  This phenomenon has been widely studied, as it offers a way to probe local PNG through the clustering of galaxies and other tracers in galaxy surveys \cite{DePutter2017_I, McCarthy2022, Krolewski2024}. 

In addition, local PNG is also expected to influence the abundance of DM halos. Numerous studies have sought to model the effects of PNG on the abundance of halos \cite{Matarrese2000, Verde2001, LoVerde2008, DAmico2011}, but comparisons with N-body simulations have revealed minor discrepancies, depending on the halo mass definitions and halo-finding methods used \cite{Grossi2007, Dalal2008, Desjacques2009}. Given these inconsistencies, it is essential to use N-body simulations to test theoretical predictions for the non-Gaussian halo mass function, and eventually develop a method to take into account the dependency on the mass definitions used to build the halo catalog.

This paper is organized as follows. In \cref{sec:png}, we provide a brief overview of the theory that describes local-type PNG and its impact on the halo mass function. In \cref{sec:compare_simulations}, we describe the suite of simulations and the data sets used in this work. We then compare theoretical predictions for the non-Gaussian halo mass function to simulation data, also analyzing the effect of PNG on halo density profiles. In \cref{sec:new_recipe}, we propose a method to encapsulate the definition of halo mass into the theoretical model for the non-Gaussian halo mass function, focusing on potential dependencies on redshift and $f_\mathrm{NL}^\mathrm{loc}$. Finally, in \cref{sec:conclusions}, we summarize the main conclusions of our work.

\section{Primordial non-Gaussianities}\label{sec:png}
\subsection{Local-type}
PNG affect the distribution of primordial perturbations that arise from inflation. Given that the magnitude of these perturbations is small, the total gravitational potential can be expressed as the sum of a Gaussian and a non-Gaussian components, the latter being characterized by the parameter $f_\mathrm{NL}$ \cite{Salopek1990, Verde2000, Komatsu2001}. In our analysis, we focus on one of the most extensively studied forms of PNG, known as the \emph{local-type}, where the gravitational potential of perturbations can be expressed as:
\begin{equation}
    \Phi(\mathbf{x}) = \phi_\mathrm{G}(\mathbf{x}) + f_\mathrm{NL}^\mathrm{loc}\left(\phi_\mathrm{G}^2(\mathbf{x})-\left\langle\phi_\mathrm{G}^2(\mathbf{x}) \right\rangle\right)\,,
    \label{eq:grav_field}
\end{equation}
where $\mathbf{x}$ is the coordinate vector and $\phi_\mathrm{G}$ represents the real-space Gaussian potential. When $f_\mathrm{NL}^\mathrm{loc}$ (hereafter $f_\mathrm{NL}$) differs from zero, the potential $\Phi$ becomes a random field characterized by a non-Gaussian probability distribution. Consequently, it cannot be fully described by the power spectrum $P_\Phi(k) = A_\mathrm{s}k^{n_\mathrm{s}-4}$ alone and higher-order moments must also be considered. Here, $A_\mathrm{s}$ and $n_\mathrm{s}$ represent the amplitude and the scalar spectral index and $k$ is the wavenumber associated to a given scale. The most significant of these higher-order contributions are \emph{bispectrum} $B_\Phi(k_1, k_2, k_3)$ and \emph{trispectrum} $T_\Phi(k_1, k_2, k_3, k_4)$.

The bispectrum and the trispectrum are respectively defined as the Fourier transform of the three-point and four-point correlation functions, and they can be expressed as:
\begin{align}
    \label{eq:def_bispectrum}
    \langle \Phi(k_1)\Phi(k_2)\Phi(k_3) \rangle & = (2\pi)^3 \delta_\mathrm{D}^{(3)}(\mathbf{k}_1 + \mathbf{k}_2 + \mathbf{k}_3) B_\Phi(k_1, k_2, k_3)\,, \\
    \langle \Phi(k_1)\Phi(k_2)\Phi(k_3)\Phi(k_4)\rangle & = (2\pi)^3 \delta_\mathrm{D}^{(3)}(\mathbf{k}_1 + \mathbf{k}_2 + \mathbf{k}_3 + \mathbf{k}_4) T_\Phi(k_1, k_2, k_3, k_4)\,,
    \label{eq:def_trispectrum}
\end{align}
where $\delta_\mathrm{D}^{(3)}$ indicates the 3D Dirac delta function. In the case of local-type PNG described by \Cref{eq:grav_field}, the bispectrum and the trispectrum reduce to the following simple forms \cite{Gong2011}:
\begin{align}
    \label{eq:bispectrum}
    B_\Phi(k_1, k_2, k_3) & = 2f_\mathrm{NL}\left[ P_\Phi(k_1)P_\Phi(k_2) +(\mathrm{2 \ cyclic}) \right]\,,\\
    T_\Phi(k_1, k_2, k_3, k_4) & = 2f_\mathrm{NL}^2\{ P_\Phi(k_1)P_\Phi(k_2)\left[ P_\Phi(k_{13}) + P_\Phi(k_{14})\right]+(\mathrm{11 \ cyclic})\}\,,
    \label{eq:trispectrum}
\end{align}
with ``cyclic'' indicating cycling permutations of the indexes. In principle, the trispectrum also includes an additional term arising from a third-order perturbation of the gravitational potential (characterized by the quantity $g_\mathrm{NL}$); however, we will disregard this complication.

It is important to recall that the gravitational potential is related to the linearly evolved density field through the following relationship:
\begin{equation}
    \delta(k, z) = \dfrac{2c^2k^2T(k)D(z)}{3\Omega_\mathrm{m}H_0^2}\Phi(k,z)=\mathcal{M}(k,z)\Phi(k,z)\,.
    \label{eq:M}
\end{equation}
Here, $c$ is the speed of light in vacuum, $T(k)$ is the transfer function of perturbations, $D(z)$ is the normalized linear growth factor, $H_0$ denotes the Hubble constant and $\Omega_\mathrm{m}$ is the present value of the matter density parameter. 

Note that in the latter expression $\delta$ refers to the unbiased density field. In cosmological scenarios characterized by non-vanishing PNG, the tracer bias can be expressed at leading order as the sum of the linear bias and a term proportional to $f_\mathrm{NL}$ \cite{Dalal2008}. Using the subscripts ``h'' and ``m'' to denote the density fields traced by DM halos and DM particles, respectively, we can write:
\begin{equation}
    \delta_\mathrm{h}(M,k,z) = b_1(M,z)\delta_\mathrm{m}(k, z) + b_\Phi f_\mathrm{NL}\Phi = \left[ b_1(M,z)+\dfrac{f_\mathrm{NL}b_\Phi}{\mathcal{M}(k,z)}\right]\delta_\mathrm{m}(k,z)\,.
\end{equation}
Here, $b_1$ is the linear bias, $b_\Phi$ is the bias parameter associated with the primordial potential $\Phi$, and $\mathcal{M}(k,z)\propto k^2$ is defined in \Cref{eq:M} and connects the gravitational potential to the density field of matter. This relation now includes a term proportional to $k^{-2}$, which introduces what is commonly indicated as \emph{scale-dependent} bias. We refer the interested reader to \Cref{app:bias} for a more detailed examination of bias in the presence of local-type PNG.

Considering again the simplified case of unbiased tracers, we use \Cref{eq:M} to write the three-point function as 
\begin{equation}
    \langle\delta(R_1)\delta(R_2)\delta(R_3)\rangle = \int \dfrac{\mathrm{d}^3k_1}{(2\pi)^3}\dfrac{\mathrm{d}^3k_2}{(2\pi)^3}\dfrac{\mathrm{d}^3k_3}{(2\pi)^3}\hat{\mathcal{M}}(k_1)\hat{\mathcal{M}}(k_2)\hat{\mathcal{M}}(k_3)\langle \Phi(k_1)\Phi(k_2)\Phi(k_3) \rangle\,,
    \label{eq:3point}
\end{equation}
where $\delta(R)$ is the smoothed density perturbation and $\hat{\mathcal{M}}(k) \coloneq \mathcal{M}(k)\hat{W}(kR)$, with $\hat{W}(kR)$ being the Fourier transform of the spherical top-hat filter function:
\begin{equation}
    \hat{W}(kR)=\dfrac{3\left[\sin(kR)-kR\cos(kR)\right]}{(kR)^3}\, .
\end{equation}
Analogous formulae are valid also for higher-order correlation functions, such as the four-point function.

\subsection{Non-Gaussian models for the halo mass function}\label{sec:models_PNG}
In the past, numerous studies have explored the impact of PNG on the halo mass function, such as the those performed by Matarrese et al. \cite{Matarrese2000} and LoVerde et al. \cite{LoVerde2008} (hereafter referred to as MVJ and LMSV, respectively). Both studies derived expressions for the non-Gaussian mass function using the Press \& Schechter approach \cite{PressSchechter1974} and including non-Gaussian terms. It is important to emphasize that the main quantity of interest is the correction factor \(\mathcal{R}(M,z)\), defined as the ratio between the non-Gaussian mass function \(n^\mathrm{NG}_\mathrm{PS}(M,z)\) and the Gaussian mass function \(n_\mathrm{PS}(M,z)\) (with the subscript ``PS'' indicating a Press \& Schechter-like expression). 

In MVJ, the authors employed the saddle-point approximation to calculate the level excursion probability, while LMSV approximated the probability density function using the Edgeworth expansion and then used it to estimate the level excursion probability. The resulting correction factors are written as follows:
\begin{align}
    \label{eq:mvj}
    \mathcal{R}_\mathrm{MVJ} (M,z) & = \exp\left(\delta_\mathrm{c}^3\dfrac{S_3}{6\sigma_M^2}\right)\left[ \dfrac{1}{6}\dfrac{\delta_\mathrm{c}}{\sqrt{1-\delta_\mathrm{c}S_3/3}}\dfrac{\mathrm{d}S_3}{\mathrm{d}\ln{\sigma_M}}+\sqrt{1-\dfrac{\delta_\mathrm{c}S_3}{3}} \right]\,,\\
    \mathcal{R}_\mathrm{LMSV} (M,z) & = 1+\dfrac{\sigma_M^2}{6\delta_\mathrm{c}}\left[S_3\left(\dfrac{\delta_\mathrm{c}^4}{\sigma_M^4}-2\dfrac{\delta_\mathrm{c}^2}{\sigma_M^2}-1\right)+\dfrac{\mathrm{d}S_3}{\mathrm{d}\ln{\sigma_M}}\left(\dfrac{\delta_\mathrm{c}^2}{\sigma_M^2}-1\right)\right]\,.
    \label{eq:loverde}
\end{align}
Here, $\delta_\mathrm{c}(z) = \delta_\mathrm{c,0}/D(z)$ denotes the linear density threshold for halo collapse, where $\delta_\mathrm{c,0}\simeq1.686$ is its present-day value, $\sigma_M$ represents the square root of the mass variance evaluated at $z = 0$ on a scale $R$ corresponding to the mass $M$, and $S_3 \coloneq \langle \delta_R^3 \rangle/\langle \delta_R^2 \rangle^2$ is the normalized skewness of the distribution (proportional to $f_\mathrm{NL}$ from \nCref{eq:bispectrum,eq:3point}). Both expressions are truncated in a way such that they only contain terms depending on the skewness. In LMSV, the authors also provided an expression for the mass function correction that includes higher-order terms in the computation of the probability density function. This resulting model is commonly referred to as second-order, or quadratic, and it is given by:
\begin{align}
    \mathcal{R}_\mathrm{LMSVQ} (M,z) = 1 &+ \dfrac{S_3\sigma_M}{6}\left( H_3(\nu)+\dfrac{1}{\nu}\dfrac{\mathrm{d}\ln{\left(S_3\sigma_M\right)}}{\mathrm{d}\ln{\sigma_M}}H_2(\nu)\right)+\nonumber\\
    &+\dfrac{\left( S_3\sigma_M\right)^2}{72}\left(H_6(\nu) + \dfrac{2}{\nu}\dfrac{\mathrm{d}\ln{\left(S_3\sigma_M\right)}}{\mathrm{d}\ln{\sigma_M}}H_5(\nu) \right)+\nonumber\\
    &+\dfrac{S_4\sigma_M^2}{24}\left(H_4(\nu) + \dfrac{1}{\nu}\dfrac{\mathrm{d}\ln{\left( S_4\sigma_M^2\right)}}{\mathrm{d}\ln{\sigma_M}}H_3(\nu) \right)\,,
    \label{eq:loverde_quad}
\end{align}
where $\nu \coloneq \delta_\mathrm{c}/\sigma_M$, $H_n(x)$ represents the Hermite polynomial of degree $n$ and $S_4 \coloneq \langle\delta_R^4\rangle/\langle\delta_R^2\rangle^3$ denotes the normalized kurtosis of the distribution (proportional to $f_\mathrm{NL}^2$ from \Cref{eq:trispectrum}). The subscript ``Q'' is added to distinguish between the linear and the quadratic model.

In contrast, in the work of D'Amico et al. \cite{DAmico2011} the authors derived the non-Gaussian halo mass function by calculating the first-crossing rate $\mathcal{F}$ of a random walk with non-Gaussian noise in the presence of an absorbing barrier. This approach was carried out perturbatively, starting with the path-integral formalism outlined in \cite{Maggiore2010I, Maggiore2010II, Maggiore2010III}. The resulting multiplicity function entering the mass function is written as $f(\sigma_M)$ = $2\sigma_M\mathcal{F}(\sigma_M)$. In terms of the ratio $\mathcal{R}$, their result (hereafter referred to as DMNP) can be expressed as:
\begin{align}
    \mathcal{R}_\mathrm{DMNP}(M,z) = &\exp\left(\dfrac{\varepsilon_1\nu^3}{6}-\dfrac{\nu^4}{8}\left( \varepsilon_1^2-\dfrac{\varepsilon_2}{3}\right)\right)\times\nonumber\\
    &\times \left\{ 1-\dfrac{\varepsilon_1\nu}{4}\left[(4-c_1)+\dfrac{1}{\nu^2}\left(c_1-\dfrac{c_2}{4}-2\right)\right] \right\}\,.
    \label{eq:damico}
\end{align}
In this expression, the coefficients $c_1$ and $c_2$ are smoothly varying functions of the variance $\sigma_M^2$. Furthermore, we define the functions $\varepsilon_n$, known as the ``equal-time'' functions, where $\varepsilon_{n-2} \coloneq \langle\delta_R^n\rangle/\langle\delta_R^2\rangle^n$ for $n\geq3$. It is straightforward to show that the equal-time functions $\varepsilon_1$ and $\varepsilon_2$ are related to normalized skewness and kurtosis by $\varepsilon_1 = \sigma_MS_3$ and $\varepsilon_2 = \sigma_M^2S_4$.

Skewness and kurtosis, along with the equal-time functions, should be calculated for each mass value \( M \). However, the numerical integration required for these computations is time consuming. To accelerate the process, we employ empirical relations for \( S_3 \) and \( S_4 \) of the form \cite{Yokoyama2011}:

\begin{equation}
    S_3 \simeq f_\mathrm{NL}\dfrac{\alpha}{\sigma_M^{2\beta}}\,, \qquad S_4\simeq f_\mathrm{NL}^2\dfrac{\gamma}{\sigma_M^{2\theta}}\,,
    \label{eq:fitting_functions}
\end{equation}
with $\alpha=2.16\times10^{-4}$, $\beta=0.4$, $\gamma=8.43\times10^{-8}$ and $\theta=0.99$. It is important to emphasize that these empirical formulae are calibrated specifically for the cosmology used in our simulations (see \cref{sec:simulations}). Consequently, if different cosmological parameters are adopted, the empirical functions will need to be re-calibrated accordingly.

Finally, we introduce the key approximation that these correction factors are independent of the specific model used to describe the mass function. Consequently, once the correction factor $\mathcal{R}(M,z)$ is determined, it can be applied to any Gaussian model $n(M,z)$, such as those proposed in \cite{ShethTormen1999, Jenkins2001, Warren2006, Watson2013, Tinker2008}, to obtain the desired non-Gaussian model:
\begin{equation}
    n^\mathrm{NG}(M,z) = \mathcal{R}(M,z) \, n(M,z)\,.
\end{equation}

In general, all models predict that positive values of the parameter $f_\mathrm{NL}$ lead to an increase in the halo number counts at high masses and a decrease at low masses. The specific mass scale at which this change occurs varies with the model; for example, the MVJ model predicts that this turnover occurs at lower masses with respect to the other models considered in this work. In contrast, negative values of $f_\mathrm{NL}$ are expected to produce the opposite effect.

\section{Validation with N-body Simulations}\label{sec:compare_simulations}
\subsection{Numerical setup}\label{sec:simulations}
The most effective way to test the models discussed in \cref{sec:models_PNG} is by utilizing numerical N-body simulations that incorporate various levels of local-type PNG, described by specific values of the parameter $f_\mathrm{NL}$. For this purpose, we utilize the \textsc{Quijote} simulations \cite{Quijote}, a suite consisting of more than $\num{82000}$ full N-body simulations.\footnote{\url{https://quijote-simulations.readthedocs.io}} Specifically, we focus on the set of simulations that features a standard $\Lambda$ cold dark matter ($\Lambda$CDM) model, as well as the set that includes PNG, known as \textsc{Quijote-png} \cite{QuijotePNG}. These simulations were generated using the codes \texttt{2LPTIC} \cite{Crocce2006} and \texttt{2LPTPNG} \cite{Scoccimarro2012} to produce the initial conditions at $z=127$ and \texttt{GADGET-3} (an enhanced version of \texttt{GADGET-2} \cite{Springel2005}) to follow their evolution up to $z=0$. For the purposes of our analysis, we utilize $8000$ realizations for the $\Lambda\mathrm{CDM}$ cosmology and all $1000$ realizations featuring local PNG ($500$ for $f_\mathrm{NL}=100$ and $500$ for $f_\mathrm{NL}=-100$). Each realization follows the evolution of $512^3$ CDM particles in a periodic box with size $L_\mathrm{box}=1 \ h^{-1}\ \mathrm{Gpc}$, which translates into a mass resolution of $6.56\times10^{11} \ h^{-1}\ \mathrm{M}_\odot$. Both sets of simulations are characterized by the same cosmological parameters: $\Omega_\mathrm{m}=0.3175$, $\Omega_\mathrm{b}=0.049$, $h=0.6711$, $n_\mathrm{s}=0.9624$, and $\sigma_8=0.834$. The only difference is the parameter $f_\mathrm{NL}$, which is set to $100$ and $-100$ in the simulations that include PNG. With the exception of $f_\mathrm{NL}$, the cosmological parameters are consistent with the constraints of \emph{Planck} 2015 \cite{Planck2015_XIII}. The main characteristics of the simulation suites are summarized in \Cref{tab:sims_prop}.
 
In addition to these publicly available suites, we ran another set of complementary simulations, maintaining a setup as similar as possible to that used in \textsc{Quijote}. Specifically, these new N-body simulations, referred to as \textsc{Pringls} (PRImordial Non-Gaussianity of Local-type Simulations), share most of the characteristics of the previously mentioned simulations, including cosmological parameters. The main differences lie in the code used to generate the initial conditions, the number of realizations, and the values of $f_\mathrm{NL}$. Local-type PNG are introduced in the initial conditions using the \texttt{PNGRUN} code \cite{Wagner2010}, with $10$ realizations for each cosmological scenario. Furthermore, we include two additional values of $f_\mathrm{NL}$ ($-40$ and $40$) compared to \textsc{Quijote-png}. We also report in \Cref{tab:sims_prop} the properties of these simulations.
\begin{table}[t]
    \centering
    \resizebox{\textwidth}{!}{
    \begin{tabular}{ccccccc}
        \toprule\toprule
        & IC & $f_\mathrm{NL}$ & $N_\mathrm{real}$ & $L_\mathrm{box}\ [h^{-1} \ \mathrm{Mpc}]$ & $N_\mathrm{part}$ & $M_\mathrm{DM}\ [h^{-1} \ \mathrm{M}_\odot]$\\
        \midrule
        \vspace{1mm}
        \textsc{Quijote} & \texttt{2LPTIC} & 0 & 8000 & 1000 & $512^3$ &$6.56\times10^{11}$\\
        \vspace{1mm}
        \textsc{Quijote-png} & \texttt{2LPTPNG} & $\pm100$ & $2\times500$ & = & = & = \\
        \textsc{Pringls} & \texttt{PNGRUN} & $0,\pm40,\pm100$ & $5\times10$ & = & = & = \\ 
        \bottomrule\bottomrule
    \end{tabular}
    }
    \caption{\emph{From left to right}: Main characteristics of the N-body simulation sets employed in this work: the code used to generate the initial conditions, the values of $f_\mathrm{NL}$, the number of realizations, their box size, the number of particles, and the DM particle mass.}
    \label{tab:sims_prop}
\end{table}

Our analysis relies on the usage of DM halo catalogs, which were obtained using the publicly available code \texttt{ROCKSTAR} \cite{Rockstar}, which refines \emph{Friends-of-Friends} (FOF) groups using an adaptive hierarchical approach across six phase-space dimensions and one time dimension.\footnote{A Friends-of-Friends algorithm \cite{Davis1985} classifies halos as groups of DM particles that are closer to each other than a specified linking length, $\ell=b\,\Bar{d}$. Here, $\Bar{d}$ represents the average inter-particle separation, and $b$ is a configurable parameter of the code.} For determining halo masses, \texttt{ROCKSTAR} calculates spherical overdensities using various user-specified density thresholds, $\Delta$. These thresholds can include the virial threshold \cite{Bryan1998} or overdensities relative to the critical or background density, $\rho_\mathrm{c} \equiv 3 H^2(z) / (8 \pi G)$ and $\rho_\mathrm{b} \equiv \rho_\mathrm{c}(z) \, \Omega_\mathrm{m}(z)$ respectively. For \textsc{Quijote} and \textsc{Quijote-png} simulations, we have used the publicly available catalogs \cite{Coulton2023, Jung2023, Jung2023_II} at $z=0,\ 0.5$ and 1, in which masses are characterized by $\Delta_\mathrm{b}=200$, and $\Delta_\mathrm{c}=200$, 500, and 2500, where the subscripts ``b'' and ``c'' denote overdensities relative to $\rho_\mathrm{b}$ and $\rho_\mathrm{c}$, respectively. These catalogs were realized assuming a parameter $b=0.27$ for the linking length of FOF groups and considering halos as collections of at least $10$ particles. For \textsc{Pringls}, we generated catalogs using the same setup as for \textsc{Quijote}, defining halo masses according to $\Delta_\mathrm{b}=200$, and $\Delta_\mathrm{c}=200$, 300, 500, 800, 1000 and 1200 at the same redshifts considered for \textsc{Quijote}.

All cosmological calculations, Bayesian analyses and data manipulations described in this work are performed in the numerical environment provided by the \texttt{CosmoBolognaLib} \cite{Marulli2016}, a large set of \emph{free software} C++/Python libraries composed of constantly growing and improving codes for cosmological analyses.\footnote{\url{https://gitlab.com/federicomarulli/CosmoBolognaLib}} Moreover, we also rely on the Python libraries offered by \texttt{Pylians} \cite{Pylians} to analyze the numerous snapshots of the simulations employed in this work. 

\subsection{Data-theory comparison}\label{sec:datavstheory}
\begin{figure}[t]
    \centering
    \includegraphics[width=0.57\linewidth]{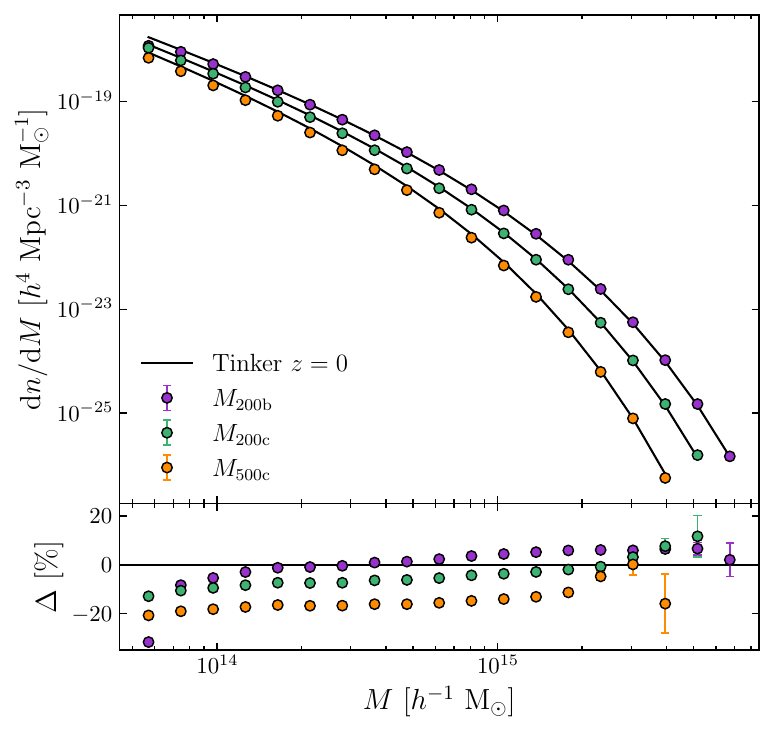}
    \caption{The average Gaussian halo mass function measured from 8000 realizations of the \textsc{Quijote} simulations at redshift $z=0$, within the mass range of $5\times10^{13}$ to $7\times10^{15}$ $h^{-1}\ \mathrm{M}_\odot$. The purple, green and orange circles correspond to different mass definitions: $M_\mathrm{200b}$, $M_\mathrm{200c}$, and $M_\mathrm{500c}$, respectively. The solid lines indicate the phenomenological model proposed by Tinker et al. \cite{Tinker2008} for each definition of mass. The subpanel displays the percentage residuals of the data relative to the model.}
    \label{fig:gauss_hmf}
\end{figure}

We begin this section by examining the Gaussian mass function derived from the standard $\Lambda \mathrm{CDM}$ simulations. \cref{fig:gauss_hmf} presents the Gaussian halo mass function as measured in the \textsc{Quijote} simulations, focusing on \texttt{ROCKSTAR} halos with masses defined by $M_\mathrm{200b}$, $M_\mathrm{200c}$, and $M_\mathrm{500c}$ at redshift $z=0$. For each definition, the mass function is calculated as the average of the 8000 individual mass functions obtained from each realization. The error for each mass bin is estimated as $\sigma/\sqrt{N_\mathrm{real}}$, where $\sigma$ represents the standard deviation between all realizations, and $N_\mathrm{real}$ is the total number of realizations. Given the large number of realizations in \textsc{Quijote}, the statistical error on the mean is significantly reduced compared to that resulting from the analysis of an individual halo mass function.

We also compare the estimated mass functions with the phenomenological model proposed by Tinker et al. \cite{Tinker2008}, which can be expressed as:
\begin{equation}
    \dfrac{\mathrm{d}n}{\mathrm{d}M} = f\left(\sigma_M\right)\dfrac{\Bar{\rho}_\mathrm{m}}{M}\dfrac{\mathrm{d}\ln{\sigma_M^{-1}}}{\mathrm{d}M}\,, \quad \mathrm{with} \quad f\left( \sigma_M\right) = A\left[\left( \dfrac{\sigma_M}{b} \right)^{-a} + 1 \right]\mathrm{e}^{-c/\sigma_M^2}\,.
\end{equation}
Here, $\Bar{\rho}_\mathrm{m}$ denotes the mean comoving density of matter, and $f\left(\sigma_M\right)$ represents the multiplicity function, which is parameterized by four variables: $A$, $a$, $b$ and $c$. These parameters are determined by fitting the data obtained from various simulation sets. It is important to note that Tinker et al. model is calibrated for halo masses defined by overdensities relative to the background density. To align our halo definitions with this model, we convert the $\Delta_\mathrm{c}$ values characterizing our halos into $\Delta_\mathrm{b}$ by dividing them by the matter density parameter $\Omega_\mathrm{m}(z)$. Since the density parameter varies with redshift, each value $\Delta_\mathrm{c}$ (e.g., 200, 500, 2500) corresponds to a different value $\Delta_\mathrm{b}$ at different $z$. 

As illustrated in the subplot of \cref{fig:gauss_hmf}, the model does not reproduce the data with high accuracy. For halos defined by $M_\mathrm{200b}$, the agreement is relatively good, with residuals around $5\%$ (excluding the first two data points). However, as we move to other mass definitions, the agreement deteriorates, with residuals increasing up to 10\% and 20\% (in absolute value) for $M_\mathrm{200c}$ and $M_\mathrm{500c}$, respectively. Similar results are also observed when the mass functions are examined at higher redshifts. We repeated the analysis on the halo mass functions extracted from \textsc{Pringls} and verified its consistency with the \textsc{Quijote} data in all definitions of mass and redshifts. Also in this case the phenomenological model again showed poor agreement with the \textsc{Pringls} data, despite the larger error bars due to the smaller number of available realizations (only $10$ per cosmological model).

Because of the high discrepancies found between the model and the Gaussian data, we chose to use the measured Gaussian mass function as a reference to calculate the deviation of the non-Gaussian halo mass function. In other words, we substituted the halo number counts predicted by the Tinker et al. model with the data extracted from the \textsc{Quijote} simulations, which have an associated statistical uncertainty that is almost negligible. Although this approach cannot be applied to real data, it eliminates systematic errors related to poor agreement between the data and the mass function model and allows us to directly measure the expected deviations from Gaussianity.

\begin{figure}[t]
    \centering
    \includegraphics[width=0.725\linewidth]{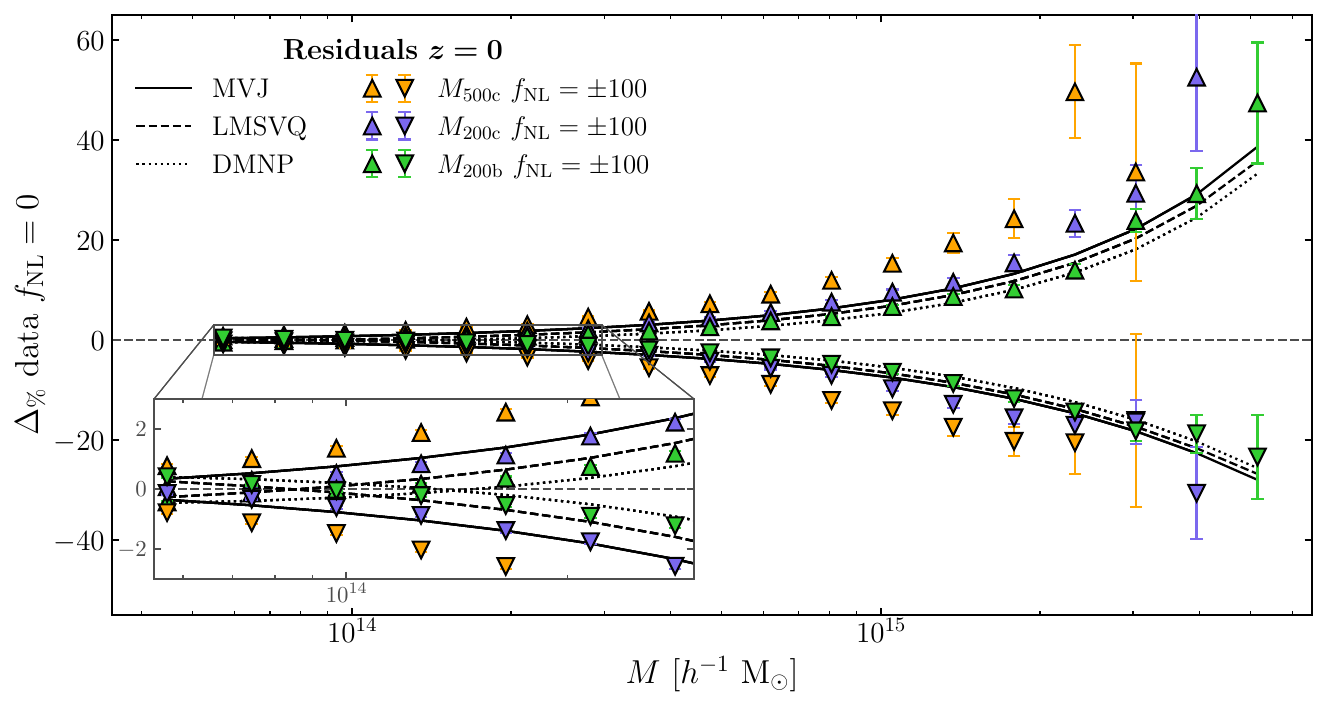}
    \caption{Deviations between halo mass functions in PNG scenarios with respect to their Gaussian counterpart. We compare different theoretical models and data sets at $z=0$, for the two scenarios $f_\mathrm{NL}=100$ and $-100$. On the $y$-axis we show the percentage residuals (\Cref{eq:residuals}) between the mean halo number counts measured using \textsc{Quijote-png} and those extracted from \textsc{Quijote} (representing the reference $\Lambda\mathrm{CDM}$ number counts). Different marker shapes are used for the two $f_\mathrm{NL}$ values: upward-pointing triangles for $100$ and downward-pointing triangles for $-100$. Colors denote different mass definitions: green for $M_\mathrm{200b}$, blue for $M_\mathrm{200c}$, and orange for $M_\mathrm{500c}$. The theoretical models for mass function corrections are indicated with different line styles: solid for MVJ, dashed for LMSVQ, and dotted for DMNP (\Cref{eq:mvj,eq:loverde_quad,eq:damico}, respectively). The subplot displays a zoomed region at low masses to allow a clearer understanding of the behavior of both model and data points.}
    \label{fig:hmf_corr_models_comparison}
\end{figure}

In this context, \cref{fig:hmf_corr_models_comparison} displays the residuals at $z=0$ between the mean halo mass functions measured in \textsc{Quijote-png} (one for $f_\mathrm{NL}=100$ and one for $f_\mathrm{NL}=-100$) and the reference Gaussian halo mass function derived from \textsc{Quijote}. The residuals are expressed as percentages through the following relation:
\begin{equation}
    \Delta_\mathrm{\%}\,(M_i) = 100\left(\dfrac{\mathrm{d}\bar{n}/\mathrm{d}M(M_i, f_\mathrm{NL}\neq0)}{\mathrm{d}\bar{n}/\mathrm{d}M(M_i,f_\mathrm{NL}=0)}-1\right)\,,
    \label{eq:residuals}
\end{equation}
where $\mathrm{d}\bar{n}/\mathrm{d}M$ represents the mean halo mass function and $M_i$ indicates its various mass bins. \cref{fig:hmf_corr_models_comparison} also shows the deviations predicted by the MVJ, LMSVQ, and DMNP models (\Cref{eq:mvj,eq:loverde_quad,eq:damico}, respectively). Theoretical models fail to accurately reproduce the data, with the agreement worsening as the overdensity used to define halo masses increases. LMSVQ and DMNP models provide a reasonable fit for $M_\mathrm{200b}$ halos, while MVJ aligns more closely with the $M_\mathrm{200c}$ case. However, all models exhibit slight over- and underpredictions in the amplitude of the deviation from the Gaussian case (the standard $\Lambda$CDM scenario) and the behavior of halos characterized by other mass definitions, such as $M_\mathrm{500c}$ and $M_\mathrm{2500c}$ --- the latter not displayed for clarity reasons --- is not well reproduced by any of the theoretical models. We also found analogous trends at $z=0.5$ and 1.

Testing the models against the data extracted from \textsc{Pringls} simulations with the same $f_\mathrm{NL}$ values led to very similar results. In fact, despite larger error bars, the observed deviations for the same mass definitions are fully consistent with those shown in \cref{fig:hmf_corr_models_comparison}, as expected.

Previous studies \cite{Grossi2009, Pillepich2010, Wagner2012} found that similar discrepancies also arise for halos identified by a FOF algorithm. In particular, both the LMSV and the MVJ corrections have been evidenced to overpredict the effect of PNG in this scenario. However, these differences could be mitigated by introducing an external factor $q$ into the calculations for the corrections of the halo mass function. This factor modifies the present linear collapse threshold as $\delta_\mathrm{c, 0} = 1.686\sqrt{q}$, with fits to numerical simulations indicating a value $q \simeq 0.75-0.8$. This adjustment is usually attributed to the effects of ellipsoidal collapse, which are reflected in a deviation from sphericity in the case of FOF halos. However, this modification does not improve the description of halos identified using a spherical overdensity algorithm, since in this case halos are spherical by construction. 

Consequently, in \cref{sec:new_recipe}, we will introduce a new parameterization that incorporates the dependency on the halo mass definition into the theoretical framework, without interpreting this correction as due to the shape of the overdensities. The following section will focus on exploring the physical motivation behind the necessity of incorporating a dependency on the density contrast threshold into the theoretical model of the mass function in PNG scenarios.

\subsection{Density profiles of dark matter halos}\label{sec:densityprofiles}

To have hints about the behavior observed in \cref{fig:hmf_corr_models_comparison}, we compute the stacked halo density profiles using \textsc{Pringls} simulations. In this case, we do not use the \textsc{Quijote-png} data sets because the total number of halos in \textsc{Pringls} is sufficient to provide good enough statistics and we are interested in exploring multiple values of $f_\mathrm{NL}$. A similar density profile analysis was performed by \cite{Smith2011} using FOF halos in simulations with $f_\mathrm{NL}=\pm100$. In contrast, we decided to focus on halos with masses $M_\mathrm{200b}$ identified by \texttt{ROCKSTAR}, extending the analysis also to smaller values of $|f_\mathrm{NL}|$. 

Density profiles are derived by counting DM particles within concentric and logarithmically spaced spheres of radii $R_i$, which range from $0.05 R_\mathrm{200b}$ to $3 R_\mathrm{200b}$. Here, $R_\mathrm{200b}$ is defined as the radius where the density is 200 times the background density. The profile estimate is then represented in terms of the dimensionless density contrast:
\begin{equation}
    \delta(R_i) = \frac{1}{\bar{n}_{\mathrm{DM}}} \frac{N(R_i)}{V(R_i)} - 1 \,,
\end{equation}
where $\Bar{n}_\mathrm{DM}$, $N(R_i)$, and $V(R_i)$ denote the mean number density of DM particles in the simulation box, the number of particles within the $i$-th sphere and its volume, respectively. This method was applied to all halos within the mass range of $ 10^{14}$ to $4 \times 10^{15} \, h^{-1} \, M_{\odot}$, using snapshots at $z = 0$ for all available realizations.

\begin{figure}[t]
    \centering
    \includegraphics[width=0.9\linewidth]{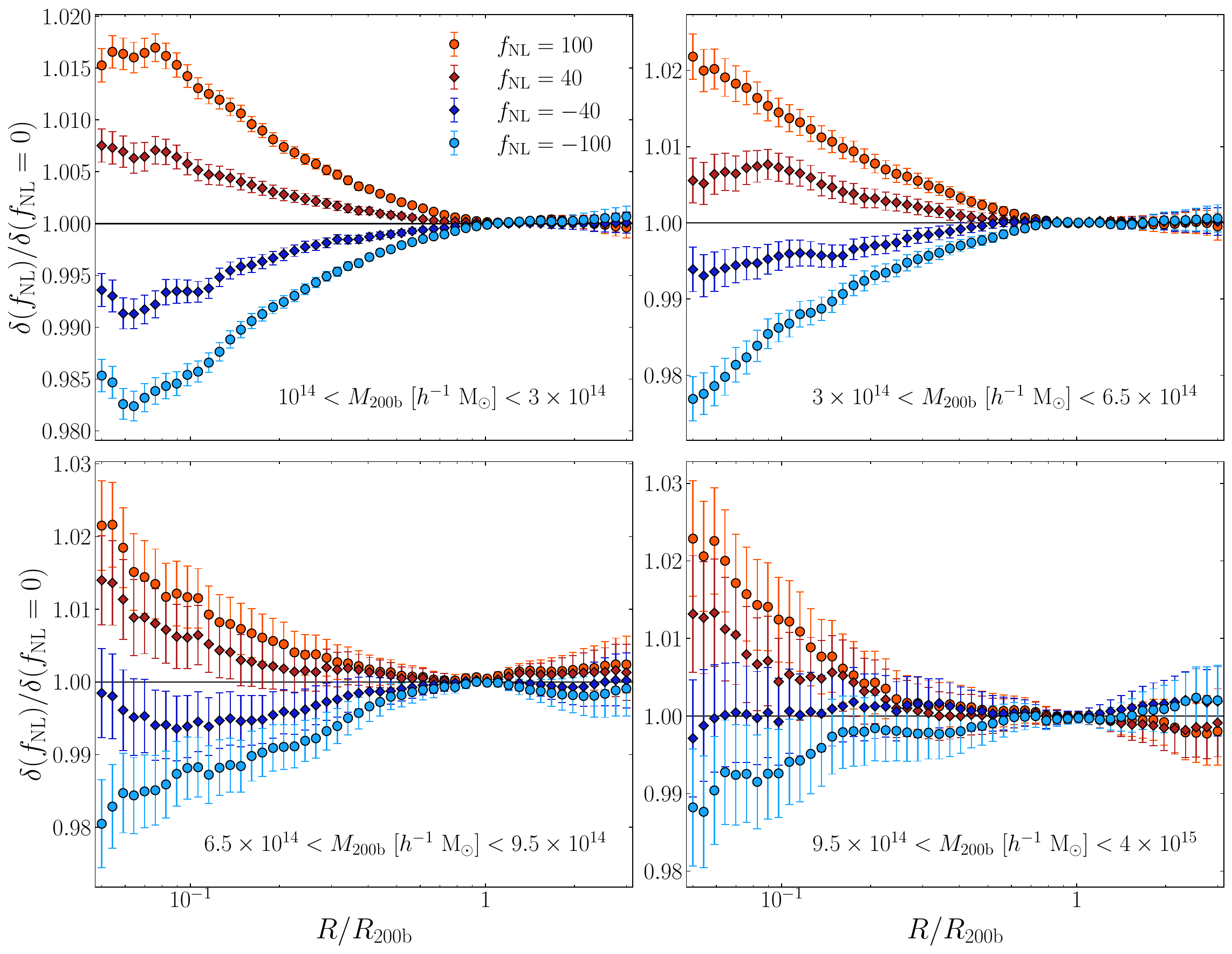}
    \caption{Ratio at $z=0$ of stacked halo density profiles from \textsc{Pringls} with different levels of PNG to those measured in the Gaussian $\Lambda\mathrm{CDM}$ scenario. In light blue, blue, red, and orange we show the results for $f_\mathrm{NL}=-100, \ -40, \ 40,$ and $100$, respectively. The panels correspond to different $M_\mathrm{200b}$ bins, covering in total the range between $10^{14}$ and $4\times10^{15} \ h^{-1}\ \mathrm{M}_\odot$.}
    \label{fig:dm_density_profile}
\end{figure}

To avoid stacking density profiles of halos having significantly different masses, we divide the individual profiles into four bins according to the mass of the halos. Finally, for each mass bin we average the individual profiles of realizations sharing the same value of $f_\mathrm{NL}$ (0, $\pm40,\ \pm100$), assigning an error to each radial bin equal to $\sigma/\sqrt{N_\mathrm{halos}}$, where $\sigma$ is the standard deviation between individual profiles and $N_\mathrm{halos}$ is the total number of halos in each mass bin. To assess deviations from Gaussianity, we normalized the mean profiles measured in simulations with $f_\mathrm{NL} \neq 0$ by profile obtained from the Gaussian $\Lambda\mathrm{CDM}$ simulations, and then propagated the error of the latter on the ratio. The results are shown in \cref{fig:dm_density_profile}.

DM halos exhibit an increase in inner density for positive values of $f_\mathrm{NL}$, while a decrease is observed when $f_\mathrm{NL}$ is negative. By construction, all profiles become identical when the \textit{x}-axis reaches unity; in fact, because of the mass definition used, every halo is characterized by the same density contrast at a distance $R=R_\mathrm{200b}$ from the center. Furthermore, the measured mean deviations from the Gaussian profile appear symmetric, with variations of at most 2\% in the innermost regions of the halos. Within each mass bin, the number of halos increases as $f_\mathrm{NL}$ increases, with the opposite behavior occurring as $f_\mathrm{NL}$ decreases, as shown in \Cref{tab:number_halos}. Interestingly, the \emph{bottom right panel} of \cref{fig:dm_density_profile} reveals that the $f_\mathrm{NL}=-40$ and $-100$ scenarios deviate less from the Gaussian case compared to their positive $f_\mathrm{NL}$ counterparts. This asymmetry suggests that the impact of negative amplitudes of PNG on the internal structure of high-mass halos is milder. However, this bin is characterized by the largest statistical uncertainty; therefore, we plan to investigate this aspect in future analyses using high-resolution simulations.

The change we observe in the slope of the density profiles, i.e. in the compactness of DM halos, is reflected in the deviations found in the halo mass function for different values of the identification threshold (see \cref{fig:hmf_corr_models_comparison}). In fact, given a threshold $\Delta$, the radius at which that density contrast is reached depends on the value of $f_\mathrm{NL}$, and the mass of the halo changes as a consequence. For example, for positive values of $f_\mathrm{NL}$, the threshold is met at a greater distance from the halo center, resulting in a larger mass within the corresponding sphere. Additionally, higher $\Delta$ values cause the identified halo radii to be closer to the center, leading to greater deviations from the Gaussian case. Notably, the asymmetry observed at large masses in the density profiles is mirrored by a corresponding asymmetry in the deviations from the Gaussian case of the HMF, further supporting the interpretation that the modifications in halo structure due to PNG directly impact the abundance of halos. Our goal is to incorporate these effects into the theoretical model of the halo mass function by adding a corrective factor to account for the variations caused by the PNG in halo density profiles.

\begin{table}[t]
    \centering
    \scalebox{1}{
    \begin{tabular}{cccccc}
    \toprule\toprule
    \multirow{2}{*}[-2pt]{$M_\mathrm{200b}\ [10^{14}\ h^{-1}\ \mathrm{M}_\odot]$} & \multicolumn{5}{c}{Number of halos}\\
    \cmidrule{2-6}
    & $f_\mathrm{NL}=100$ & $f_\mathrm{NL}=40$ & $f_\mathrm{NL}=0$ & $f_\mathrm{NL}=-40$ & $f_\mathrm{NL}=-100$\\
    \midrule
    \vspace{1mm}
    $1-3$ & \num{285787} & \num{285216} & \num{284678} & \num{284332} & \num{283663}\\
    \vspace{1mm}
    $3-6.5$ & \num{48059} & \num{47471} & \num{47088} & \num{46580} & \num{45983} \\
    \vspace{1mm}
    $6.5-9.5$ & \num{7085} & \num{6913} & \num{6775} & \num{6666} & \num{6488} \\
    $9.5-40$ & \num{3795} & \num{3602} & \num{3492} & \num{3375} & \num{3206}\\
    \bottomrule\bottomrule
    \end{tabular}
    }
    \caption{Total number of DM halos across each set of 10 \textsc{Pringls} simulations characterized by a specific $f_\mathrm{NL}$ value at $z=0$: from left to right $100,\ 40, \ 0, \ -40$ and $-100$. Each row corresponds to a specific mass range, with the mass defined by the overdensity $\Delta_\mathrm{b}=200$.}
    \label{tab:number_halos}
\end{table}

\section{A New Parameterization for the Mass Function Correction}\label{sec:new_recipe}
\subsection[Calibration of the correction factor \texorpdfstring{$\kappa$}{}]{Calibration of the correction factor \texorpdfstring{\boldmath{$\kappa$}}{}}\label{sec:kappa_calibration}
Based on the results presented in the previous sections, it is clear that the theoretical models, without adjustments for specific mass definitions, do not adequately agree with data from simulations, especially at higher overdensities. To address this, we sought a method that incorporates the effects of varying mass definitions. In the literature, it is common to introduce a correction factor to align theoretical models with FOF data sets when correcting the halo mass function for PNG (as discussed in \cref{sec:datavstheory}). We adopted a similar approach by introducing a modification of the linear critical density for collapse, parameterized by a factor $\kappa$, which varies with the chosen overdensity criterion used to identify the halos. Among all the theoretical non-Gaussian mass function models analyzed, we chose to rely on the LMSVQ model since it showed the best agreement with our data (see e.g. the $M_\mathrm{200b}$ case in \cref{fig:hmf_corr_models_comparison}). Consequently, we modify \Cref{eq:loverde_quad} by substituting $\delta_\mathrm{c}$ with $\kappa \times \delta_\mathrm{c}$. Therefore, thanks to this additional factor, our model for the non-Gaussian HMF more closely resembles the structure of commonly used models for the Gaussian HMF, where the quantity $\delta_\mathrm{c}$ is typically multiplied by or replaced with a fitted parameter \cite[see, e.g.,][]{ShethTormen1999, Jenkins2001, Warren2006, Watson2013, Tinker2008}.

We start by focusing on \textsc{Quijote-png} data sets, as the high number of realizations available for these simulations is expected to provide a more accurate and precise calibration of this factor. In order to extend the LMSVQ model to halos with any mass definitions, we perform a Bayesian analysis using a Markov chain Monte Carlo (MCMC) algorithm to estimate the correction factor required for data sets created with different overdensity thresholds. Since we are interested in modeling the deviations from the Gaussian mass function, our data are given by the residuals between the halo number counts extracted from the simulations with $f_\mathrm{NL}\neq0$ and those with $f_\mathrm{NL}=0$, as defined in \Cref{eq:residuals}. For this fit we set a flat prior on the only free parameter of the model (i.e. $\kappa$) and a Gaussian likelihood function expressed as:
\begin{equation}
    \mathcal{L}\left(\{x_i\} | \{\sigma_i\}, \{\mu_i\}\right)=\prod_{i=1}^N \dfrac{1}{\sqrt{2\pi\sigma_i^2}}\exp\left( -\dfrac{\left(x_i-\mu_i\right)^2}{2\sigma_i^2} \right)\,,
    \label{eq:likelihood}
\end{equation}
where $N$ is the total number of data points, and $x_i$, $\sigma_i$, and $\mu_i$ are the data, the error, and the model at the mass bin $i$, respectively. We repeat the procedure to cover all the available redshifts and for both cosmological scenarios $f_\mathrm{NL}= \pm 100$. \Cref{tab:kappa} shows the resulting values of $\kappa$ derived for different redshifts and mass definitions. For a fixed mass definition, we found no significant differences between the fits of the $f_\mathrm{NL}=100$ and $-100$ data sets, as almost all estimated parameters are consistent within the uncertainties. This suggests that $\kappa$ is independent of the sign of $f_\mathrm{NL}$ (see \cref{sec:kappa_fnl} for further details). In contrast, the values of $\kappa$ derived for halos defined by a critical overdensity threshold, $\Delta_\mathrm{c}$, exhibit a slight dependence on redshift, with $\kappa$ decreasing as redshift increases.

\begin{table}[t]
    \centering
    \resizebox{\textwidth}{!}{
    \begin{tabular}{ccccccc}
    \toprule\toprule
    \multirow{2}{*}[-2pt]{Mass} & \multicolumn{2}{c}{$\kappa(z=0)$} & \multicolumn{2}{c}{$\kappa(z=0.5)$} & \multicolumn{2}{c}{$\kappa(z=1)$}\\
    \cmidrule{2-7}
    & $f_\mathrm{NL}=100$ & $f_\mathrm{NL}=-100$ & $f_\mathrm{NL}=100$ & $f_\mathrm{NL}=-100$ & $f_\mathrm{NL}=100$ & $f_\mathrm{NL}=-100$\\
    \midrule
    \vspace{1mm}
    $M_\mathrm{200b}$ & \small{$0.971 \pm 0.003$} & \small{$0.976 \pm 0.003$} & \small{$0.978 \pm 0.002$} & \small{$0.980 \pm 0.002$} & \small{$0.982 \pm 0.002$} & \small{$0.979 \pm 0.002$} \\
    \vspace{1mm}
    $M_\mathrm{200c}$ & \small{$1.091 \pm 0.003$} & \small{$1.097 \pm 0.003$} & \small{$1.021 \pm 0.002$} & \small{$1.022 \pm 0.002$} & \small{$0.998 \pm 0.002$} & \small{$0.995 \pm 0.002$} \\
    \vspace{1mm}
    $M_\mathrm{500c}$ & \small{$1.243 \pm 0.003$} & \small{$1.245 \pm 0.003$} & \small{$1.139 \pm 0.002$} & \small{$1.136 \pm 0.003$} & \small{$1.089 \pm 0.002$} & \small{$1.081 \pm 0.002$}\\
    $M_\mathrm{2500c}$ & \small{$1.670 \pm 0.004$} & \small{$1.665 \pm 0.004$} & \small{$1.509 \pm 0.004$} & \small{$1.508 \pm 0.004$} & \small{$1.415 \pm 0.006$} & \small{$1.396 \pm 0.007$}\\
    \bottomrule\bottomrule
    \end{tabular}
    }
    \caption{The correction factor $\kappa$ as a function of redshift and mass definition for the two $f_\mathrm{NL}$ values from the \textsc{Quijote-png} simulations. Parameter values and their associated errors are estimated through MCMC fitting.}
    \label{tab:kappa}
\end{table}

To find a relation that can be used for halos identified at any redshift, we convert all critical overdensities $\Delta_\mathrm{c}$ ($200$, $500$, and $2500$) into background overdensities $\Delta_\mathrm{b}$ by dividing by $\Omega_\mathrm{m}(z)$. In our case, each value of $\Delta_\mathrm{c}$ corresponds to three values of $\Delta_\mathrm{b}$, one for each redshift of $0$, $0.5$ and $1$. The advantage of converting these thresholds in terms of background densities lies also in removing the dependence on $h$: given the current uncertainty on the estimation of the Hubble constant, it is important to calibrate a relation that is independent of the value of $h$ used to build the cosmological simulations.\footnote{Although current measurements suggest a value of $H_0 \simeq 70 \ \mathrm{km \ s^{-1} \ Mpc^{-1}}$, the precise determination of the Hubble constant is still debated due to discrepancies between the results derived with low- and high- redshift probes \cite{DiValentino2021}. For a review of the complications associated with quantities that depend explicitly on $h$ see, e.g., \cite{Sanchez2020}.}

\begin{figure}[t]
    \centering
    \includegraphics[width=0.45\linewidth]{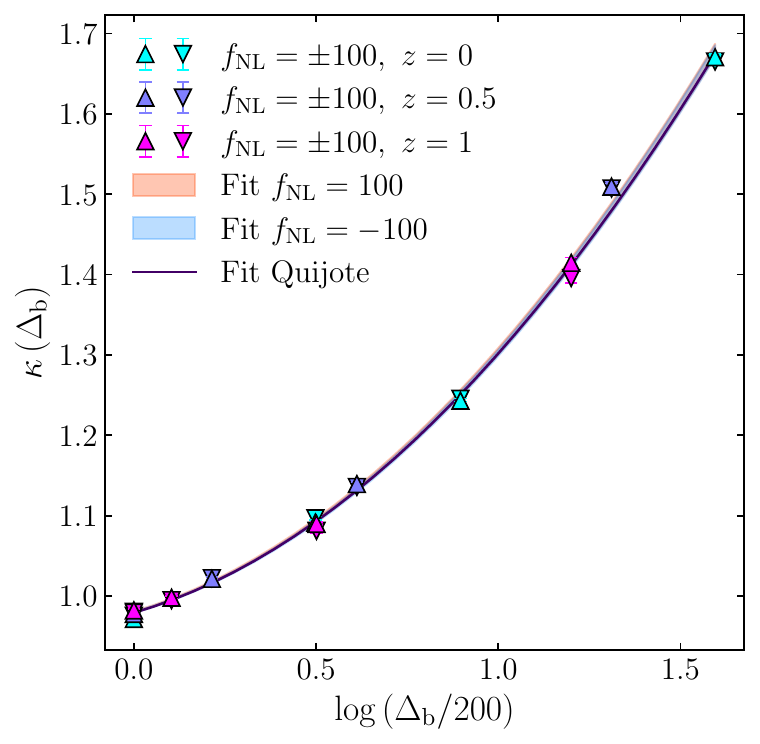}
    \hspace{0.4cm}
    \includegraphics[width=0.46\linewidth]{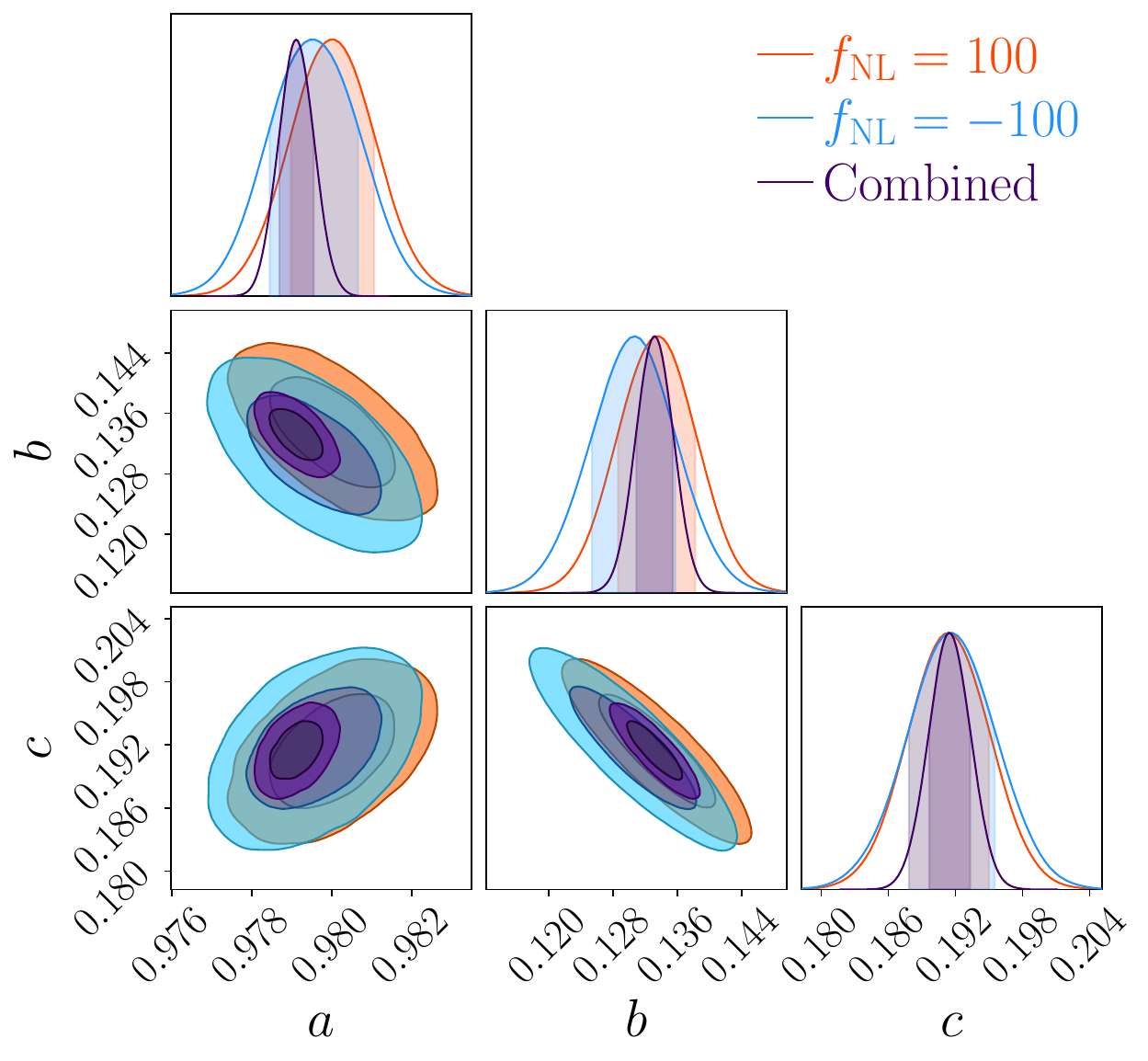}
    \caption{\emph{Left}: factor $\kappa$ as a function of the overdensity $\Delta_\mathrm{b}$. Different markers indicate the two opposite values of the $f_\mathrm{NL}$ parameter in the \textsc{Quijote-png} simulations (upward-pointing triangles for $100$ and downward-pointing triangles for $-100$), while colors correspond to redshifts $z=0$, $0.5$, and $1$ (cyan, light purple, and magenta, respectively). In red and blue we indicate the $1\sigma$ confidence region around the best fit of the $f_\mathrm{NL}=100$ and $-100$ data points, respectively (note that these regions are almost perfectly overlapping). The dark purple curve represents the simultaneous fit of both data sets, including the effect of covariance. \emph{Right:} 68\% and 95\% 2D confidence contours for the parameters $a$, $b$, and $c$ characterizing the second-degree polynomials (\Cref{eq:kappa_nocov}) for each fitting procedure. The same colors are used to indicate the results for $f_\mathrm{NL}=100$, $-100$ and their combination. The projected 1D marginalized posterior distributions are shown at the top of each column along with the 68\% uncertainty (shaded bars).}
    \label{fig:factor_kappa}
\end{figure}

Now our goal is to search for a relation between the different values of the factor $\kappa$ and the corresponding background density contrasts used to define the halos. For this purpose, we rely on a second-degree polynomial of the form:
\begin{equation}
    \kappa(x) = a + bx + cx^2\, \quad \text{with} \quad x=\log(\Delta_\mathrm{b}/200)\,,
    \label{eq:kappa_nocov}
\end{equation}
and fit the coefficients $a$, $b$ and $c$ for the two scenarios $f_\mathrm{NL} = \pm 100$. The best-fit values of these coefficients are $a=0.980\pm0.001 \ (0.979\pm0.001)$, $b=0.134\pm0.005 \ (0.131\pm0.005)$, $c=0.191\pm0.004 \ (0.191\pm0.004)$ for $f_\mathrm{NL}=100 \ (-100)$.
The results of this fit are shown in \cref{fig:factor_kappa}. In the \emph{left panel} all the values of $\kappa$ are displayed as a function of $\Delta_\mathrm{b}$. From this plot, we can appreciate how the data points closely follow the shape of the second-degree polynomial and how the results for $f_\mathrm{NL}=100$ and $f_\mathrm{NL}=-100$ are highly compatible. In the \emph{right panel} of \cref{fig:factor_kappa} we instead show the posterior distributions of the polynomial coefficients derived from the fit of $\kappa (\Delta_\mathrm{b})$. As a further confirmation, here we can see that the confidence contours for $f_\mathrm{NL}=\pm100$ cosmologies are consistent within $1\sigma$. We conclude therefore that our re-parameterization of the halo mass function model can be considered independent of the sign of $f_\mathrm{NL}$.

Given the latter result, a simple improvement of the analysis just presented consists in modeling simultaneously the deviations measured for $f_\mathrm{NL}=100$ and $f_\mathrm{NL}=-100$ to increase the precision in the calibration of the parameters $a$, $b$ and $c$. 
However, since realizations with opposite sign of $f_\mathrm{NL}$ share the same random seed for initial conditions, these data sets cannot be considered as independent. Therefore, to follow this strategy we need to consider the covariance between the mass bins in the two different cosmological scenarios and across all mass definitions used. 

\begin{figure}[t]
    \centering
    \includegraphics[width=\linewidth]{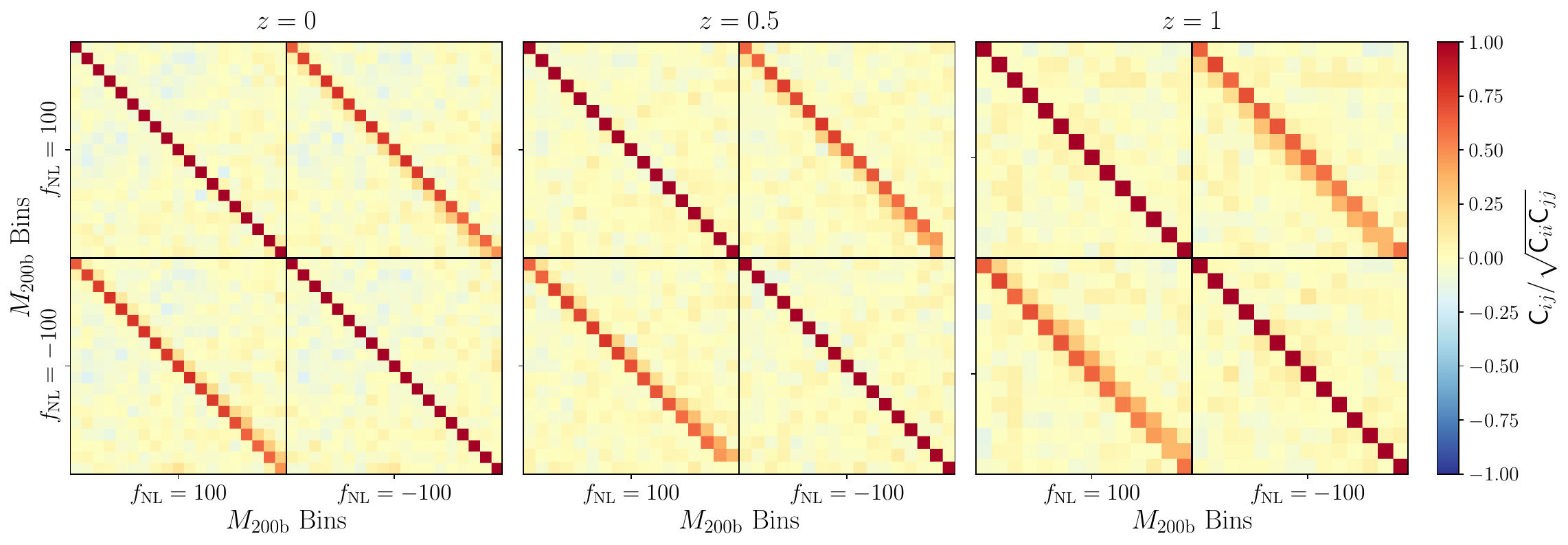}
    \caption{Covariance matrices (normalized by their diagonal elements) of the residuals between the $2\times500$ individual halo mass functions measured in \textsc{Quijote-png} for $f_\mathrm{NL}=100$ and $-100$ and the mean one measured in the \textsc{Quijote} standard $\Lambda\mathrm{CDM}$ simulations. Each subplot corresponds to the $M_\mathrm{200b}$ covariance matrix at different redshifts: 0 (\emph{left}), 0.5 (\emph{center}), and 1 (\emph{right}). Each matrix is divided into four blocks separated by vertical and horizontal black lines to isolate the two different cosmological scenarios.}
    \label{fig:cov_matrix}
\end{figure}

\cref{fig:cov_matrix} presents the covariance matrix \textsf{C} normalized by the diagonal elements, estimated from the residuals between the individual $2\times500$ halo mass functions $M_\mathrm{200b}$ measured in each realization of \textsc{Quijote-png} ($500$ for $f_\mathrm{NL}=100$ and $500$ for $f_\mathrm{NL}=-100$) and the mean halo mass function extracted from \textsc{Quijote} ($f_\mathrm{NL}=0$). Similar covariance matrices are also obtained for the other mass definitions. Each element of these matrices is rescaled by a factor of $500$ to account for the number of realizations; in this way, the diagonal elements will correctly represent the error on the mean of the residuals. \cref{fig:cov_matrix} also shows a strong correlation between the same mass bins in the $f_\mathrm{NL}=\pm100$ data sets, whereas for other bins, the correlation is consistent with zero, except for minor fluctuations due to noise. To avoid possible inaccuracies due to the latter, we set to zero all elements whose correlation coefficient lies between $-0.15$ and $0.15$.
The resulting covariance matrices were then used as input for the MCMC analysis to constrain $\kappa$ for each redshift and mass definition, generalizing \Cref{eq:likelihood} as follows:
\begin{equation}
    \mathcal{L}\left( \mathbf{x} | \textsf{C}, \boldsymbol{\mu}\right)=\dfrac{1}{\left(2\pi\right)^{N/2}\det(\textsf{C})^{1/2}}\exp\left( -\dfrac{1}{2}(\mathbf{x}-\boldsymbol{\mu})^{\mathrm{T}}\textsf{C}^{-1}(\mathbf{x}-\boldsymbol{\mu})\right)\,,
\end{equation}
where $N$ is the total number of data points, $\mathbf{x}$ is the data vector, and $\boldsymbol{\mu}$ is the vector of expected values. The results for this case are also shown in \cref{fig:factor_kappa}. As expected, the new best fit of $\kappa(x)$ is consistent with that obtained from the analysis of the individual cases $f_\mathrm{NL}=100$ and $f_\mathrm{NL}=-100$. This is evident in the \emph{right panel}, where the posterior distribution of the second-degree polynomial parameters is shown to be consistent with those of the earlier cases. The best fit we obtained for this case is:
\begin{equation}
    \kappa(x) = 0.9792 + 0.132x + 0.190x^2\,,
    \label{eq:kappa_cov}
\end{equation}
with $x$ defined in \Cref{eq:kappa_nocov}. The errors associated with each polynomial parameter are $\Delta a = 0.0004$, $\Delta b = 0.002$, and $\Delta c = 0.002$. Consequently, we observe a reduction in the errors of approximately $50\%$ compared to those obtained from individual fits.

\begin{figure}[p]
    \centering
    \includegraphics[width=0.78\linewidth]{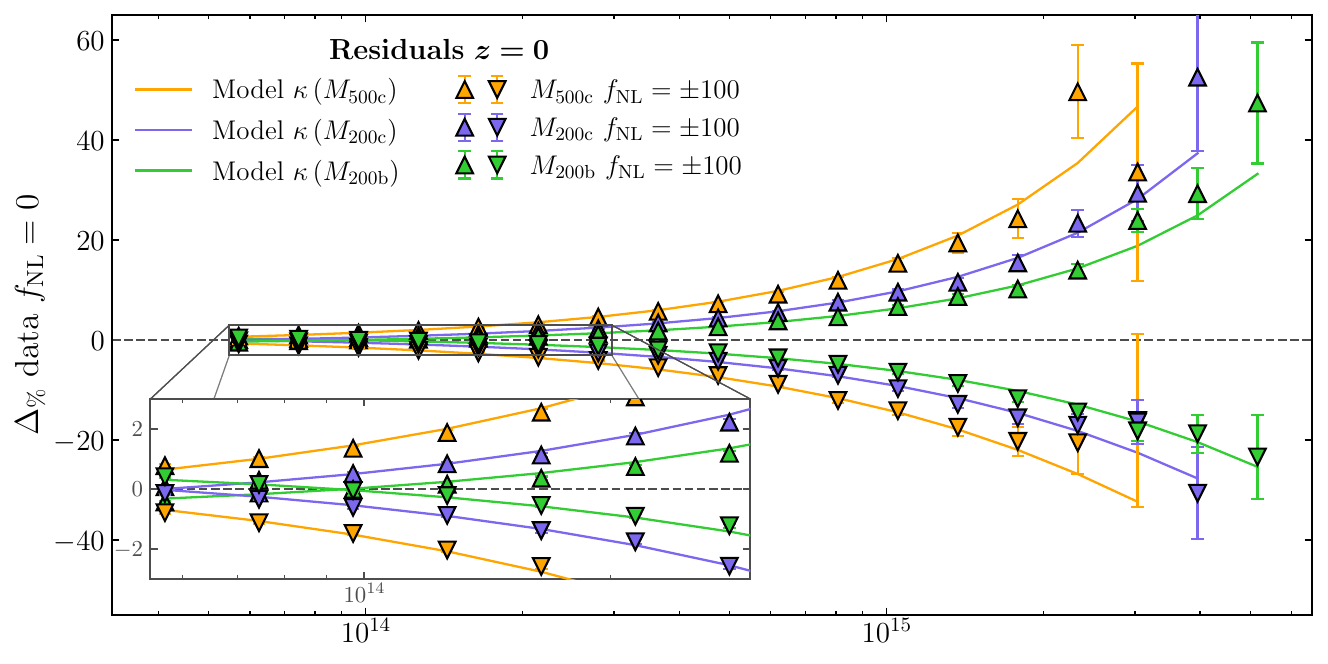}
    \includegraphics[width=0.78\linewidth]{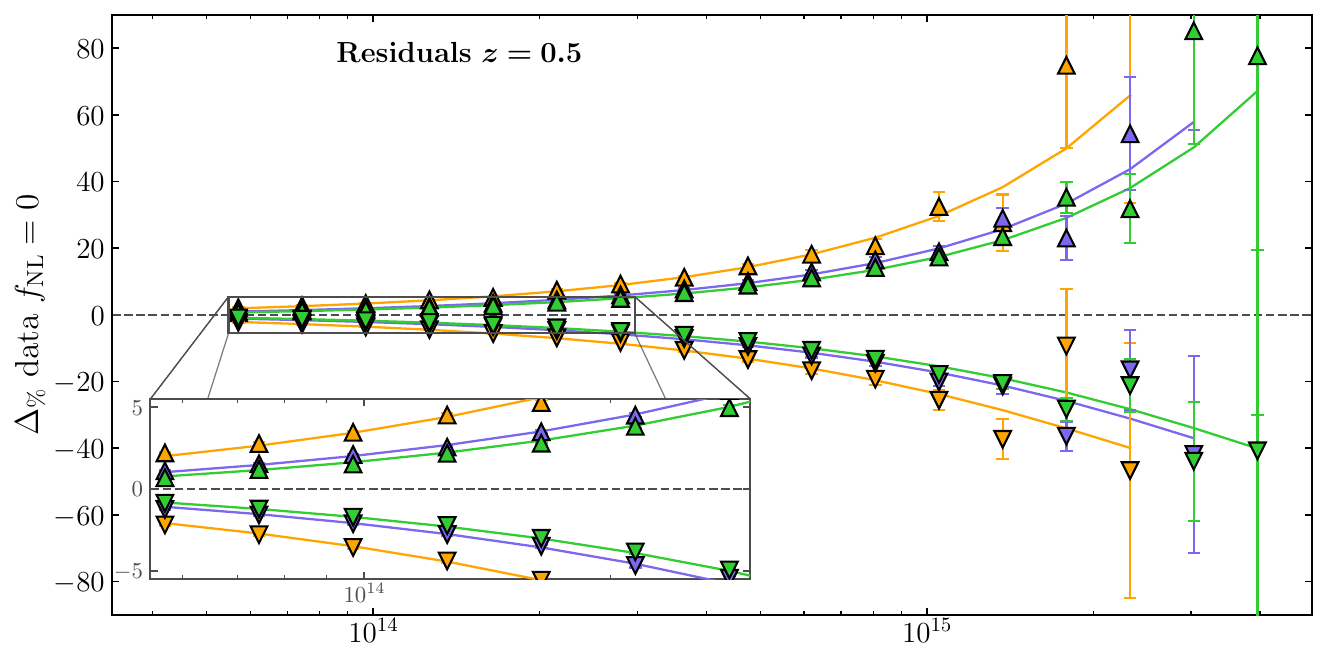}
    \includegraphics[width=0.78\linewidth]{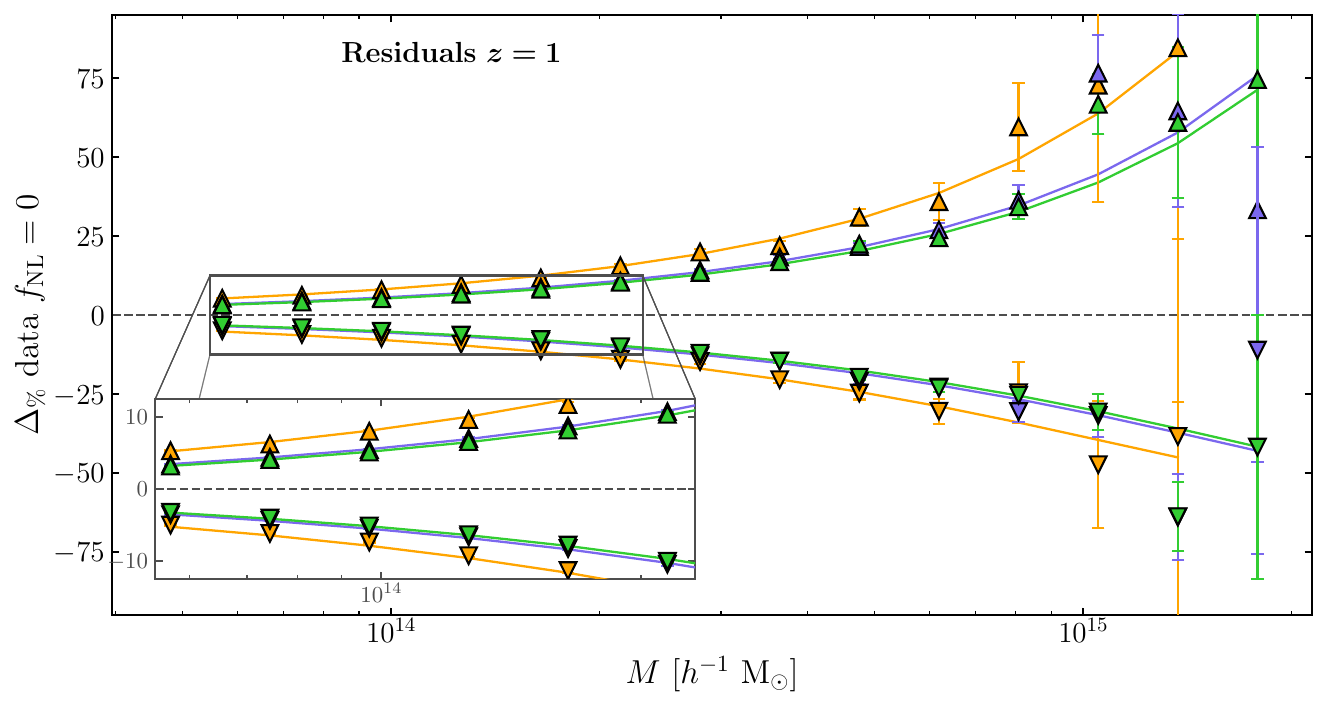}
    \caption{\emph{From top to bottom}: residuals in percentage (\Cref{eq:residuals}) of the halo mass function correction as a function of mass for various mass definitions across different redshifts. Markers indicate opposite $f_\mathrm{NL}$ values, with upward-pointing triangles for $f_\mathrm{NL}=100$ and downward-pointing triangles for $f_\mathrm{NL}=-100$. Green, blue and orange refers to the mass definitions $M_\mathrm{200b}$, $M_\mathrm{200c}$, $M_\mathrm{500c}$, respectively.  Finally, with solid lines we indicate the LMSVQ model corrected by the $\kappa$ factor. The colors of the lines matches the colors of the corresponding mass definition. In each panel the subplot displays a zoomed region at low masses to allow a clearer understanding of the behaviour of both model and data points.}
    \label{fig:new_recipe_hmf_correction}
\end{figure}

Finally, \cref{fig:new_recipe_hmf_correction} compares the data with the new parameterization for the LMSVQ model across redshifts and mass definitions $M_\mathrm{200b}$, $M_\mathrm{200c}$, and $M_\mathrm{500c}$. With the addition of the factor $\kappa\left(\Delta_\mathrm{b}\right)$, the model accurately reproduces the data in all considered cases (including $M_\mathrm{2500c}$). Notably, as shown in the \emph{top panel}, now we accurately capture the point at low masses where the effect of $f_\mathrm{NL}$ on halo counts reverses, which was not well constrained by the original implementation. 

To evaluate the goodness of this fit, we calculate the reduced chi-square, $\Tilde{\chi}^2$, for the residuals of the two cosmological scenarios $f_\mathrm{NL}=100$ and $-100$. When considering the LMSVQ model reparameterized by means of $\kappa$, the agreement with the data is remarkable, as $\Tilde{\chi}^2$ is typically of order $1-2$. For example, at $z=0$ and for $f_\mathrm{NL}=100$ we observe $\Tilde{\chi}^2=1.4,\ 2.8$ and $1.4$ for the definitions of mass $M_\mathrm{200c},\ M_\mathrm{500c}$ and $M_\mathrm{2500c}$, respectively. The improvement is particularly significant, as $\Tilde{\chi}^2$ evaluated with the original model is $46$, $233$ and $773$, respectively for the same mass definitions. Similar results are also obtained at higher redshifts and for the $f_\mathrm{NL}=-100$ case.

\subsection[Testing different \texorpdfstring{$f_\mathrm{NL}$ values}{}]{Testing different \texorpdfstring{\boldmath{$f_\mathrm{NL}$ values}}{}}\label{sec:kappa_fnl}
In the previous section, we demonstrated how the correction factor $\kappa$ appears to be independent of the sign of $f_\mathrm{NL}$. To further test the non-Gaussian halo mass function re-parameterization, we compare it with the data extracted from \textsc{Pringls}, which also includes scenarios with $f_\mathrm{NL} = \pm 40$. Thus, the same analysis performed with \textsc{Quijote-png} is now repeated using the halos identified in \textsc{Pringls}. The \textsc{Pringls} sets with $f_\mathrm{NL}=\pm100$ will be used to verify the consistency with the  results obtained with \textsc{Quijote-png}.

Now, because of the limited number of realizations, we expect the derived constraints to be less precise. Additionally, the reduced number of realizations prevents us from simultaneously modeling halo number counts with different values of $f_\mathrm{NL}$: in this case the covariance matrix is dominated by noise and does not allow us to accurately estimate the correlation between the mass bins of different sets of simulations. Consequently, we will continue testing the calibration of $\kappa$ using individual fits as in \cref{sec:new_recipe}.

\begin{figure}[t]
    \centering
    \includegraphics[width=0.6
    \linewidth]{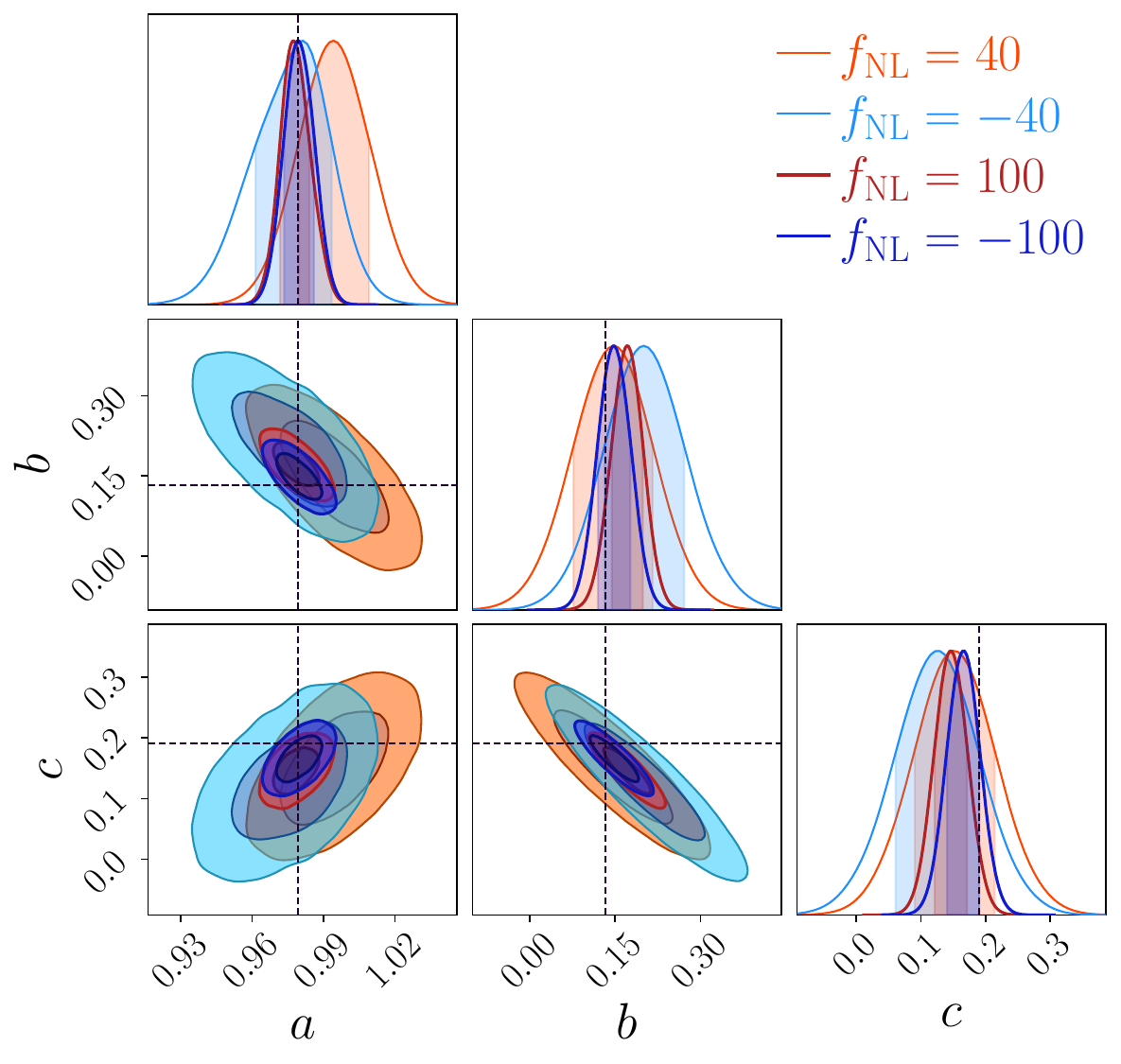}
    \caption{$68\%$ and $95\%$ confidence levels for the coefficient $a$, $b$, and $c$ of the second-degree polynomial used to fit the values of $\kappa (\Delta_\mathrm{b})$ derived with \textsc{Pringls} simulations. Blue, light blue, orange and red correspond to the different values of $f_\mathrm{NL}$: $-100$, $-40$, $40$, $100$, respectively. Black dashed lines are the best fit values of the parameters reported in \Cref{eq:kappa_cov}, which we obtained using \textsc{Quijote-png}. The projected 1D marginalized posterior distributions are shown at the top of each column  along with the
    68\% uncertainty (shaded bars).}
    \label{fig:factor_kappa2}
\end{figure}

Also in this case, we measure the mass function of halos identified with \texttt{ROCKSTAR}, transforming the thresholds $\Delta_\mathrm{c}$ (200, 300, 500, 800 1000 and 1200) in their corresponding value $\Delta_\mathrm{b}$ at $z=0$, $0,5$ and 1. Then, for each threshold value we find the correction factor $\kappa$ required to match the measured deviations from the Gaussian halo mass function (\Cref{eq:residuals}). Consequently, we fit the coefficients $a$, $b$, and $c$ of the second-degree polynomial used to parameterize $\kappa$ as a function of $\Delta_\mathrm{b}$ (see \Cref{eq:kappa_nocov}).

We present the main outcome of this analysis in \cref{fig:factor_kappa2}, where we show the confidence contours relative to $a$, $b$ and $c$ for the four scenarios $f_\mathrm{NL} = \pm40$ and $f_\mathrm{NL} = \pm100$. We note that the posterior distributions of these parameters are in good agreement with the best-fit values obtained with \textsc{Quijote-png} (vertical and horizontal dashed lines in the plot, also reported in \Cref{eq:kappa_cov}). A minor deviation is observed for the $f_\mathrm{NL} = 100$ case, which is, however, compatible at the $2\sigma$ level with our reference values. We consider this minor discrepancy negligible for our analysis, attributing it to statistical fluctuations due to the presence of noise in the data. Despite using the same number of realizations, we observe in \cref{fig:factor_kappa2} that the contours for the $f_\mathrm{NL}=\pm40$ scenarios are broader. This is because for $f_\mathrm{NL}$ values closer to zero, the differences from the $\Lambda$CDM case are smaller, and the relatively large errors of the residuals limit the constraining power.

Most importantly, given the compatibility of the results obtained for all the scenarios analyzed, we can conclude that the calibrated relation for $\kappa(\Delta_\mathrm{b})$ is not only independent of the sign of $f_\mathrm{NL}$, but also of its absolute value. We attribute this result to the fact that, in the expression for the halo mass function correction, all the information on the magnitude of $f_\mathrm{NL}$ is already well encapsulated in $S_3$ and $S_4$. Having no statistical evidence to the contrary, we conclude that it is not necessary to recalibrate the parameters of the Gaussian mass function according to the value of $f_\mathrm{NL}$. This outcome is clearly very important: given a sample of DM halos identified with any threshold $\Delta$, we now have a unique and simple recipe to predict the mass function in scenarios featuring whatever level of local-type PNG.

Furthermore, we note that the proposed re-parametrization of $\delta_\mathrm{c}$ by means of the factor $\kappa(\Delta_\mathrm{b})$ can be applied also to other non-Gaussian halo mass function models, for example MVJ \cite{Matarrese2000} and DMNP \cite{DAmico2011}. We confirm that this modification is indeed effective in correcting the model predictions according to the selected halo density thresholds, although the best agreement with the data is achieved using the LSMVQ model. Since we used the latter to calibrate the $\kappa(\Delta_\mathrm{b})$ relation, we recommend applying \Cref{eq:kappa_cov} solely in combination with the LMSVQ model for an accurate prediction of the non-Gaussian halo mass function.

As a final consideration, we emphasize that, given the resolution of the simulations, it is not possible to totally exclude potential resolution-related effects, particularly on the estimate of the $\kappa$ parameter for halos identified with particularly high thresholds (such as $\Delta_\mathrm{c}=1200$ or greater). We refer the reader to \Cref{app:pringls_hr} for a more in-depth analysis of this subject.

\subsection{Impact on forecasted constraints}\label{sec:forecasts}
Equipped with a new prescription to correct the non-Gaussian halo mass function model for different halo identification thresholds, we now investigate its impact in forecasting constraints on $f_\mathrm{NL}$. We set $f_\mathrm{NL}$ as a free parameter with uniform prior and we compare the outcomes of two models: the original LMSVQ halo mass function correction (\Cref{eq:loverde_quad}) and its re-parametrized version, in which the halo linear collapse threshold is multiplied by the function $\kappa (\Delta_\mathrm{b})$ expressed in \Cref{eq:kappa_cov}. 

With the goal of building up a more realistic setting, we use a subset of \textsc{Quijote-png} simulations that cover a volume of $50\ h^{-3}\ \mathrm{Gpc}^3$, similar to those anticipated for upcoming surveys. In practice, we use the mean halo number counts in the mass range from $5\times10^{13}$ to  $7\times10^{15}$ $h^{-1}\ \mathrm{M}_\odot$ derived from the $50$ realizations of \textsc{Quijote-png} characterized by $f_\mathrm{NL} = 100$, and we employ the mean Gaussian halo mass function derived from \textsc{Quijote} to compute the residuals as in \Cref{eq:residuals}. We then perform a MCMC analysis, to determine the best-fit value of the parameter $f_\mathrm{NL}$ and the corresponding uncertainty. The analysis is repeated for different halo mass definitions ($M_{200\mathrm{b}}$, $M_{200\mathrm{c}}$, $M_{500\mathrm{c}}$, $M_{2500\mathrm{c}}$) and redshifts ($0$, $0.5$, $1$) to check for possible systematic errors.

\begin{table}[t]
    \centering
    \begin{tabular}{ccccccccc}
        \toprule\toprule
        \multirow{2}{*}[-2pt]{Mass} & & \multicolumn{3}{c}{$f_\mathrm{NL}$ from LMSVQ} & & \multicolumn{3}{c}{$f_\mathrm{NL}$ from LMSVQ with $\kappa(\Delta_\mathrm{b})$}\\
        \cmidrule{3-5} \cmidrule{7-9}
        & & $z=0$ & $z=0.5$ & $z=1$ & & $z=0$ & $z=0.5$ & $z=1$ \\
        \midrule
        \vspace{2mm}
        $M_\mathrm{200b}$ & & $91.5^{+4.4}_{-4.0}$ & $90.8^{+2.7}_{-3.0}$ & $92.1^{+2.1}_{-1.8}$ & & $105.1^{+4.9}_{-4.7}$ & $101.2^{+3.2}_{-3.2}$ & $98.9^{+2.3}_{-1.8}$ \\
        \vspace{2mm}
        $M_\mathrm{200c}$ & & $142.6^{+6.5}_{-7.1}$ & $111.6^{+3.4}_{-2.9}$ & $98.1^{+2.1}_{-1.9}$ & & $99.7^{+4.2}_{-4.5}$ & $101.2^{+2.7}_{-3.0}$ & $98.9^{+2.2}_{-1.9}$ \\
        \vspace{2mm}
        $M_\mathrm{500c}$ & & $237.3^{+11.8}_{-11.5}$ & $182.4.8^{+4.9}_{-5.5}$ & $139.8^{+3.4}_{-3.5}$ & & $99.9^{+3.6}_{-3.9}$ & $98.5^{+2.6}_{-3.0}$ & $98.9^{+2.7}_{-2.0}$ \\
        $M_\mathrm{2500c}$ & & $305.0^{+33.9}_{-34.8}$ & $633.1^{+24.8}_{-25.3}$ & $364.1^{+22.5}_{-16.3}$ & & $100.4^{+3.2}_{-3.2}$ & $96.7^{+3.4}_{-3.1}$ & $94.7^{+4.4}_{-5.0}$ \\
        \bottomrule\bottomrule
    \end{tabular}
    
    \caption{Constraints on $f_\mathrm{NL}$ obtained by fitting the mean non-Gaussian mass functions extracted from $50$ realizations of the \textsc{Quijote-png} simulations with $f_\mathrm{NL}=100$, for different halo mass definitions and redshifts. The left part of the table shows the values obtained using the original LMSVQ correction (\Cref{eq:loverde_quad}), while the right part refers to its re-parametrized version, that includes the dependency on the halo mass definition trough the previously calibrated factor $\kappa(\Delta_\mathrm{b})$.}
    \label{tab:fnl_constraints}
\end{table}

The best-fit values of $f_\mathrm{NL}$ and their $1\sigma$ uncertainty are presented in \Cref{tab:fnl_constraints}. For the sake of clarity, we remind the reader that these results should be compared with the true cosmological parameters of the simulations, which feature $f_\mathrm{NL}=100$. We can notice that employing the original LMSVQ model leads to biased constraints, with a deviation from the true value that increases with the threshold used to identify the halos. This offset is present at all the analyzed redshifts with a similar intensity. As expected from the results presented in \cref{sec:new_recipe}, the case with $M_{200\mathrm{b}}$ is the least affected by systematic errors.

In contrast, including $\kappa (\Delta_\mathrm{b})$ in the model significantly improves the accuracy of the constraints, which are almost always consistent with the true simulation value at the $1\sigma$ level. Minor systematic errors emerge only for extreme mass definitions (i.e. $M_{2500\mathrm{c}}$) and at $z \geq 0.5$. This result provides further confirmation that, without accounting for the halo identification threshold in the halo mass function correction for PNG, significant deviations from the theory arise. It is therefore essential to incorporate this dependency on the halo definition into the model in order to avoid biased cosmological constraints.

However, we stress that these constraints on $f_\mathrm{NL}$ are optimistic, as we have fixed all other cosmological parameters. There could be some degeneracies between the parameters characterizing our new version of the model and the cosmological ones. We aim to investigate this aspect in future analyses.

\section{Discussions and Conclusions}\label{sec:conclusions}
In this paper, we focused on testing the accuracy of already existing halo mass function models developed for cosmological scenarios featuring primordial non-Gaussianities (PNG). We selected three well-known models from the literature, that is, Matarrese et al. (2000) \cite{Matarrese2000}, LoVerde et al. (2008) \cite{LoVerde2008}, and D’Amico et al. (2011) \cite{DAmico2011}, to test against simulations. These models provide a theoretical prescription to predict the deviation from the Gaussian halo mass function produced by local type of PNG, quantified by the parameter $f_\mathrm{NL}$. The data were extracted from halo catalogs generated by applying the algorithm \texttt{ROCKSTAR} \cite{Rockstar} to the three sets of N-body cosmological simulations \textsc{Quijote}, \textsc{Quijote-png} and \textsc{Pringls} (see \cref{sec:simulations}). These simulations share the same standard $\Lambda$CDM cosmological parameters and together cover different levels of PNG, i.e. $f_\mathrm{NL}=0$,  $f_\mathrm{NL}=\pm40$ and $f_\mathrm{NL}=\pm100$. 
A key point in our analysis consisted in the creation of catalogs with different thresholds $\Delta$ for halo identification. In particular, we focused on different threshold values, defined both with respect to $\rho_\mathrm{b}$ (e.g. $\Delta_\mathrm{b}=200$) and $\rho_\mathrm{c}$ (e.g. $\Delta_\mathrm{c}=200$, $500$, $2500$).

In \cref{sec:datavstheory}, we compared the halo number counts extracted from these catalogs with the predictions of the three models we considered, finding that none of them was able to perfectly capture the deviations caused by PNG on the halo mass function. We showed how this discrepancy depends on the halo mass definition, as it increases significantly when higher density thresholds are used. To investigate the reason for this systematic error, in \cref{sec:densityprofiles} we compared the stacked density profiles of DM halos measured in simulations with different values of $f_\mathrm{NL}$. We showed that PNG modify the DM matter distribution around the halo center such that for $f_\mathrm{NL}>0$ halos exhibit a steeper inner density profile, whereas the opposite occurs for $f_\mathrm{NL}<0$. Moreover, the change in the slope of the halo density profile becomes more significant as the value of $|f_\mathrm{NL}|$ increases. This behavior explains why using different identification thresholds produces a deviation in the observed halo mass function. In fact, halos become more (less) compact in a $f_\mathrm{NL}>0 \ (<0)$ cosmology, and the higher the identification threshold, i.e. closer we get to the halo center, the larger (smaller) their mass will result with respect to a standard $\Lambda$CDM scenario.

At this point, the main part of our work consisted of developing a correction to the theoretical model for the non-Gaussian halo mass function that incorporates the dependence on the halo mass definition. We took as a reference the quadratic model of LoVerde et al. (2008) \cite{LoVerde2008} since it was the one showing the best agreement with the measured halo mass function (see $\Delta_\mathrm{b}=200$ data in \cref{fig:hmf_corr_models_comparison}).
The prescription we propose is simple: we correct the linear density threshold for halo collapse, $\delta_\mathrm{c}$, by means of a factor $\kappa(\Delta_\mathrm{b})$. The latter is calibrated as a function of the halo identification threshold, which we require to be expressed in terms of the background density of the Universe, $\rho_\mathrm{b}(z)$.

In \cref{sec:kappa_calibration} we calibrated this correction factor by fitting the deviations on the measured halo mass function caused by PNG at different redshifts ($z=0,$ $0.5$, $1$), and for halos identified with different thresholds (converting those overdensities relative to $\rho_\mathrm{c}$ into the corresponding background quantity).
We found that a second-degree polynomial can accurately represent the variation of $\kappa$ as a function of $\Delta_\mathrm{b}(z)$, for all analyzed redshifts. We estimated the best-fit values for the three polynomial coefficients and their relative uncertainty, finding that a unique relation can be used to incorporate the dependence of the halo identification threshold into the non-Gaussian halo mass function model. In other terms, we proposed a re-parametrization of the model that holds for cosmologies with different levels of local-type PNG, independent of the sign and the amplitude of the parameter $f_\mathrm{NL}$ (see \cref{sec:kappa_fnl}). The best fit of the function $\kappa(\Delta_\mathrm{b})$ is presented in \Cref{eq:kappa_cov}, where we combined the halo number counts extracted from simulations with $f_\mathrm{NL}=-100$ and $f_\mathrm{NL}=100$, including the covariance between these data.

Finally, in \cref{sec:forecasts}, we presented a simple example of cosmological forecasts focused on constraining $f_\mathrm{NL}$. We considered a survey volume of $50 \ h^{-3} \ \mathrm{Gpc}^3$ and compared the constraints derived with the original model and its re-parametrized version (see \Cref{tab:fnl_constraints}). We demonstrated how the original implementation leads to extreme biased cosmological constraints for halos identified with thresholds as $\Delta_\mathrm{c}=200$ or $500$, with a systematic error increasing for higher values of $\Delta$. In contrast, we found that the model re-parametrized with $\Delta_\mathrm{b}(z)$ allows us to accurately constrain $f_\mathrm{NL}$ without statistically relevant systematic errors.

It is important to emphasize that these results are particularly relevant in the perspective of the application to real data. Indeed, wide field surveys allow us to extract the number of galaxy clusters as a function of their mass, but their definition of mass can vary depending on the wavelength analyzed. For example, optical and X-ray observations generally identify clusters with masses $M_\mathrm{200c}$ and $M_\mathrm{500c}$, respectively, making it crucial to have a model that consistently adapts to any identification threshold. The re-parametrized model offers a natural solution to this need. However, a real application to survey data will require all cosmological parameters to be left free, not just $f_\mathrm{NL}$. Therefore, it is important to consider that non-Gaussian halo mass function models are computationally very expensive to calculate for different cosmologies, as the approximated relations in \Cref{eq:fitting_functions} cannot be used. For this reason, we aim to develop emulators of the non-Gaussian halo mass function to speed up model computation, making it feasible for use in MCMC analyses. 

Regarding this last point, although it is true that our results were obtained for relatively large values of $f_\mathrm{NL}$, we believe that similar improvements in terms of modeling systematics on the parameter would also be present when aiming to constrain more realistic values of $f_\mathrm{NL}$. However, in such cases the observable deviations would be significantly smaller than those considered here, and one would also need to account for the uncertainties in the number counts as well as those introduced, for instance, by the choice of the Gaussian mass function model. For this reason, we plan to investigate this aspect in future work.

Finally, we underline that the methodology proposed in this paper is not meant to be limited only to local-type PNG: a natural extension of this work will consist of testing the models present in the literature also for other shapes of the potential bispectrum (i.e. equilateral, enfolded, orthogonal). Moreover, one of our goals is to extend our analysis to the underdense counterpart of DM halos, namely cosmic voids. In fact, the same formalism used to predict the non-Gaussian halo mass function can be applied to void counts, allowing us to model the non-Gaussian void size function. These two models can ultimately be used in combination to extract the overall information from halo and void number counts. In fact, several studies have demonstrated the strong complementarity of these two statistics \citep[see e.g.][]{Bayer2021,Kreisch2021,Contarini2022,Pelliciari}, making them excellent probes for breaking key degeneracies between cosmological parameters.

\acknowledgments
LF thanks Matteo Esposito for his help with numerical libraries and the OPINAS LSS group for the contribution and useful suggestions they provided. LF also thanks the Max Planck Institute for Extraterrestrial Physics for its hospitality and acknowledges the financial contribution from the grant ``Borse di studio per la preparazione all'estero della tesi di laurea magistrale" of the University of Bologna. FM and LM acknowledge the financial contribution from the grant
PRIN-MUR 2022 20227RNLY3 ``The concordance cosmological model: stress-tests with galaxy clusters” supported by Next Generation EU and from the grant ASI n. 2024-10-HH.0 ``Attività scientifiche per la missione Euclid – fase E". LF, SC, FM, MB, and LM acknowledge the use of computational resources from the parallel computing cluster of the Open Physics Hub (\url{https://site.unibo.it/openphysicshub/en}) at the Department of Physics and Astronomy in Bologna. We acknowledge the use of Python libraries \texttt{NumPy} \cite{harris2020}, \texttt{Matplotlib} \cite{Hunter2007}, \texttt{Emcee} \cite{Foreman-Mackey2013}, and \texttt{ChainConsumer} \cite{Hinton2016}.

\appendix
\section{Non-Gaussian Halo Bias}\label{app:bias}
As already introduced in \cref{sec:png}, the tracer bias becomes scale dependent in cosmological scenarios characterized by non-vanishing PNG.
Consequently, the power spectrum $P_\mathrm{h}(k,z)$ of dark matter halos is related to the matter power spectrum $P_\mathrm{m}(k,z)$ by the following relation:
\begin{equation}
    P_\mathrm{h}(k,z) = \left[ b_1(M,z)+\dfrac{f_\mathrm{NL}b_\Phi}{\mathcal{M}(k,z)}\right]^2 P_\mathrm{m}(k,z)\,.
    \label{eq:P_hh_ng}
\end{equation}
The parameter $b_\Phi$ represents the response of halo abundance to perturbations in the primordial potential. In the context of peak-background split theory, it can also be interpreted as the response of the halo number density to a rescaling of the amplitude of the primordial power spectrum \cite{Barreira2022c}:
\begin{equation}
    b_\Phi = \dfrac{1}{\Bar{n}_\mathrm{h}}\dfrac{\partial \Bar{n}_\mathrm{h} }{\partial \log(f_\mathrm{NL}\Phi)} = \dfrac{1}{\Bar{n}_\mathrm{h}} \dfrac{\partial \Bar{n}_\mathrm{h}}{\partial \log(A_\mathrm{s})}\,,
\end{equation}
where $\Bar{n}_\mathrm{h}$ represents the mean number of halos, and $A_\mathrm{s}$ denotes the amplitude of the primordial scalar perturbations. Following \cite{Slosar2008}, we can express $b_\Phi$ as:
\begin{equation} \label{eq:NGnbias}
    b_\Phi = 2\delta_\mathrm{c,0}(b_1 - p)\,,
\end{equation}

where $\delta_\mathrm{c,0} = 1.686$ and $p$ is a parameter introduced to take into account possible deviations caused by the non-universality of the halo mass function model. In particular, the assumption of a universal mass function leads to $p = 1$. This expression for $p$ is often referred to as the universal relation for $b_\Phi$. However, recent studies suggest that this relation does not always accurately describe the scale-dependent bias. For example, \cite{Barreira2020} found that for galaxies selected by stellar mass, $p = 0.5$ provides a better fit for the $b_\Phi$ relation. Similarly, in \cite{Gutierrez2024}, an analysis of halos within a specific mass range found $p = 0.955 \pm 0.013$, indicating a discrepancy of about $3\sigma$ from the universal relation for $b_\Phi$. Moreover, $b_\Phi$ may depend on additional properties beyond mass, such as halo concentration \cite{Lazeyras2023, Fondi2024}.

To test the non-Gaussian bias model we use $100$ realizations for each cosmological scenario in the \textsc{Quijote} and \textsc{Quijote-png} simulations ($f_\mathrm{NL} = 0$ and $\pm100$). We restrict the analysis to halos with masses greater than $5 \times 10^{13}\ h^{-1}\ \mathrm{M}_\odot$, regardless of the mass definition used. In each realization of the three sets of simulations, we measure the halo bias by taking the ratio of the halo-matter power spectrum $P_\mathrm{h,m}(k,z)$ to the matter power spectrum $P_\mathrm{m}(k,z)$ in the $k$ interval $[6, \, 30] \times10^{-3} \ h\ \mathrm{Mpc^{-1}}$.\footnote{We verified that this bias estimate is consistent with the one given by $[P_\mathrm{h}(k,z)/P_\mathrm{m}(k,z)]^{1/2}$, although we discarded the latter since it was more affected by shot noise.} Due to the dependence $k^{-2}$ on the non-Gaussian bias, it is crucial to measure the deviation from the linear bias $\Delta b(k, f_\mathrm{NL})$, on large scales, where the signal is expected to be stronger.

\begin{figure}[t]
    \centering
    \includegraphics[width=\linewidth]{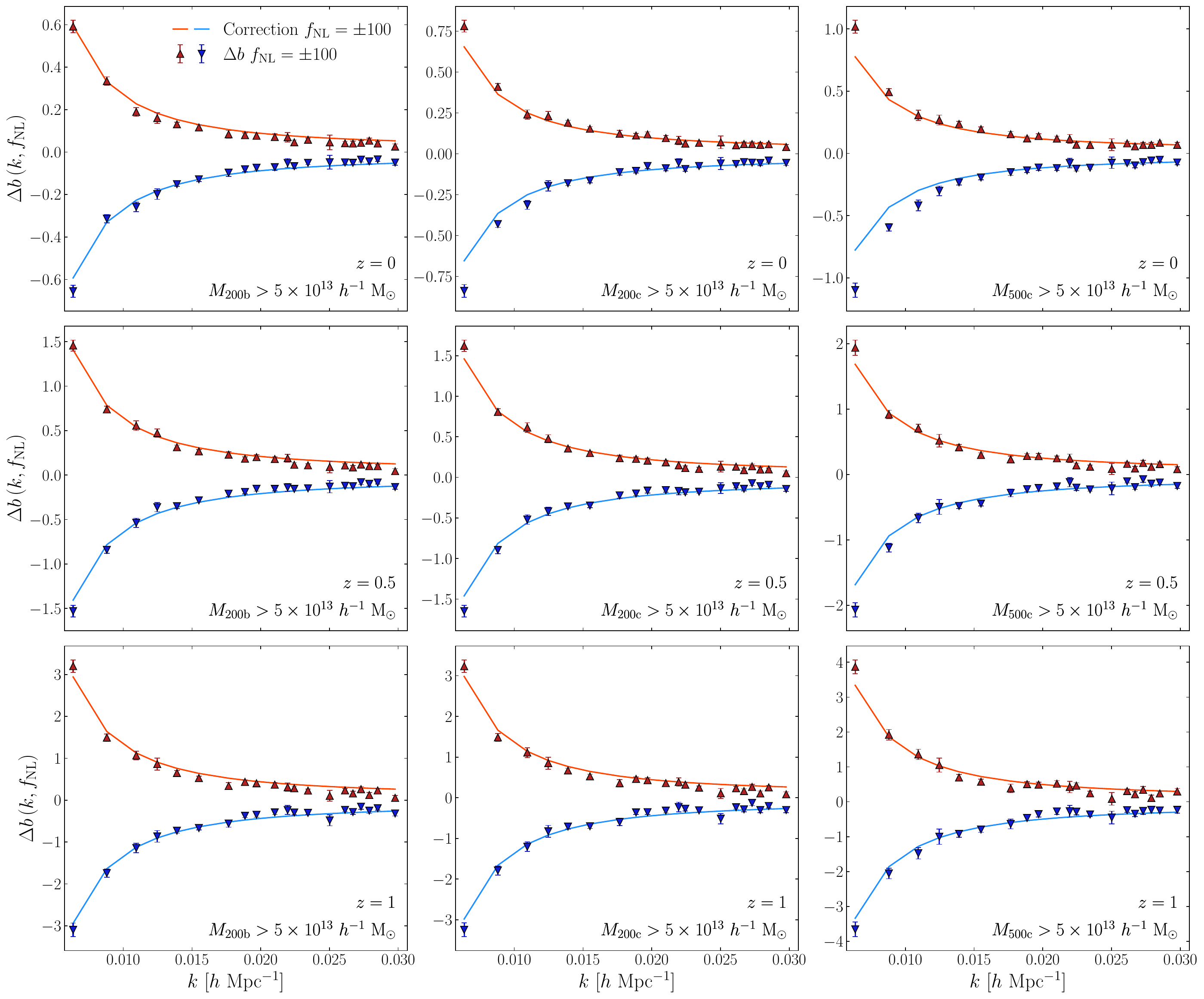}
    \caption{Deviations of the halo bias from $b_1$ caused by the presence of PNG. $\Delta b$ is the difference between the non-Gaussian and the Gaussian bias and is reported as a function of scale $k$, for different $z$ and mass definitions. Rows correspond to data sets with the same redshift (from top to bottom $z=0$, $0.5$, and $1$), while columns refer to different halo mass definitions (from left to right $M_\mathrm{200b}$, $M_\mathrm{200c}$, $M_\mathrm{500c}$). All the halos have masses above $5 \times 10^{13}\ h^{-1} \ \mathrm{M}_\odot$, regardless of the chosen halo identification threshold. The red upward-pointing triangles and the blue downward-pointing triangles represent the data extracted from simulations with $f_\mathrm{NL} = 100$ and $f_\mathrm{NL} = -100$, respectively. The orange and light blue solid lines are instead the predictions computed for the non-Gaussian halo bias assuming $p = 1$ in \Cref{eq:NGnbias}, again for $f_\mathrm{NL} = 100$ and $f_\mathrm{NL} = -100$ respectively.}
    \label{fig:ng_bias}
\end{figure}

Therefore, we average the scale-dependent bias measured in the $100$ realizations for each scenario and assign an error to each $k$ bin, calculated as the standard deviation divided by the square root of the number of realizations employed. Finally, we compute $\Delta b$ by subtracting the mean bias measured in PNG simulations by the one measured in the Gaussian $\Lambda$CDM simulations, propagating the error.
The results of this analysis are presented in \cref{fig:ng_bias}, which shows a direct comparison of the measured $\Delta b$ with the predictions from \Cref{eq:P_hh_ng}, assuming the universal relation (i.e. $p=1$). 

The plot displays the measured deviations at different redshifts and for various mass definitions. 
We observe that the overall trend of the data is well reproduced by the non-Gaussian bias model, despite the fact that small discrepancies are observed for $M_\mathrm{200b}$ halos at $z = 0.5$ and $z = 1$. We tested the effect of leaving $p$ as a free parameter of the model to reveal possible deviations from universality. However, the additional degree of freedom appears to not significantly improve the agreement of the model with the observed trend, except in a few cases. Moreover, the best-fit values obtained for $p$ range from $0.64$ to $1.47$ without a clear pattern. Thus, we conclude that no strong evidence is found for deviations from universality, at least with the precision achieved with the data employed in our analysis.

\section{Calibration against High Resolution Simulations}\label{app:pringls_hr}
Defining halo masses within a radius enclosing a given overdensity $\Delta$ implies that the halo mass can be derived from the integral of the density profile within a fixed radius. Consequently, the mass depends on the internal density distribution of the halo and is therefore more sensitive to resolution effects, especially at high overdensities.

To evaluate the impact of resolution on our parameterization, we repeated the analysis performed with the \textsc{Pringls} suite using its higher resolution counterpart, \textsc{Pringls-hr}. This set consists of only three realizations, each characterized by $1024^3$ particles: one for the Gaussian case and two for $f_\mathrm{NL}=100$ and $f_\mathrm{NL}=-100$. Despite the increase in the number of particles, the number of realizations implies that the confidence contours will be significantly wider than those obtained from \textsc{Quijote}. We compute the halo mass functions from the halos identified with \texttt{ROCKSTAR}, adopting the same overdensity thresholds used in \textsc{Pringls}. Following the same procedure already outlined in this paper, we convert the critical overdensity threshold $\Delta_\mathrm{c}$ into the corresponding $\Delta_\mathrm{b}$ values at $z=0,\ 0.5$, and $1$, then we determine the correction factor $\kappa$ required to match the residuals across all mass definitions and redshifts. Finally, we extract through a MCMC analysis the three parameters defining the second-order polynomial used to model the behavior of $\kappa$.  

\begin{figure}[t]
    \centering
    \includegraphics[width=0.575\linewidth]{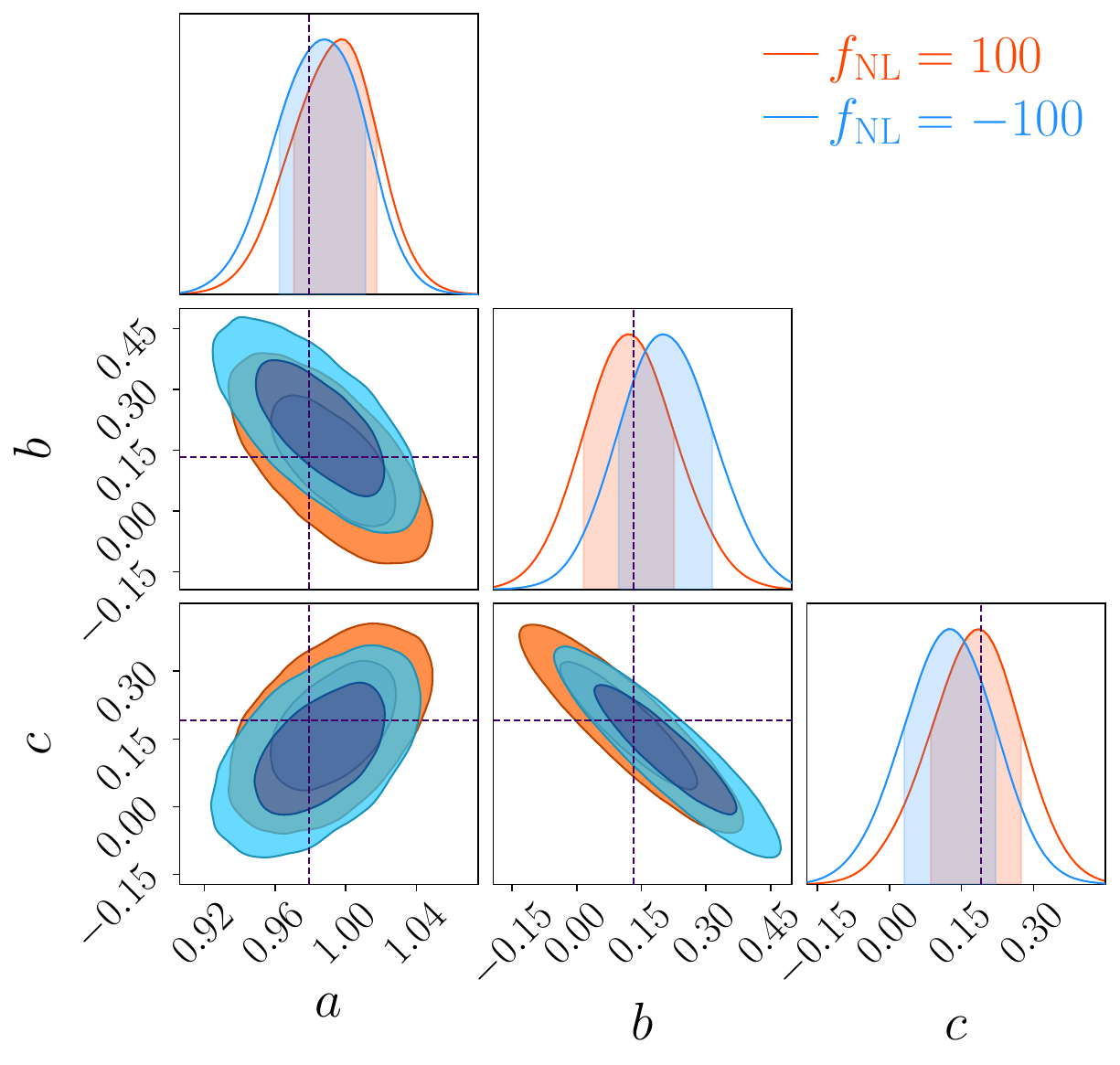}
    \caption{$68\%$ and $95\%$ confidence levels for the coefficients $a$, $b$, and $c$ of the second-degree polynomial used to fit $\kappa (\Delta_\mathrm{b})$ from \textsc{Pringls-hr} simulations. Light blue and orange contours correspond to $f_\mathrm{NL}=-100$ and $f_\mathrm{NL}=100$, respectively. Black dashed lines indicate the best-fit values of the parameters reported in \Cref{eq:kappa_cov}, obtained using \textsc{Quijote-png}. The projected 1D marginalized posterior distributions are shown at the top of each column, with shaded bars representing the 68\% uncertainty.}
    \label{fig:factor_kappa_hr}
\end{figure}

The results of this analysis are shown in \cref{fig:factor_kappa_hr}. We notice how the contours derived from the residuals for $f_\mathrm{NL}=100$ and $-100$ are fully compatible, highlighting that the independence of this new parameterization of the sign of $f_\mathrm{NL}$ is robust even when considering high-resolution simulations. Moreover, we observe from the posterior distributions that the results obtained with the \textsc{Quijote-png} data sets (vertical and horizontal dashed lines) are compatible within $1\sigma$ with those of \textsc{Pringls-hr}. Therefore, we conclude that our new parameterization does not seem to be significantly affected by resolution effects. However, further studies with additional realizations would help to confirm this prediction.

\bibliographystyle{JHEP}
\bibliography{biblio}

\end{document}